 \documentclass[pteplogo]{ptephy_v2} 

\preprintnumber{XXXX-XXXX} 
\usepackage{bm}
\usepackage{amsmath}
\usepackage{graphicx}
\usepackage{url}
\usepackage{dcolumn}
\usepackage[colorlinks=true, linkcolor=blue, citecolor=blue, urlcolor=blue]{hyperref}
\usepackage{subcaption}
\usepackage{float}
\usepackage{braket}
\usepackage{float}
\usepackage{braket}
\usepackage{threeparttable}

\begin{document}

\title{Mirror and triplet energy differences in $sd$-shell nuclei using microscopic interactions with isospin-symmetry breaking effects}

\author[1,2]{Chandan Sarma}
\affil[1]{Department of Physics, Indian Institute of Technology Roorkee, Roorkee 247667, India}	
\affil[2]{School of Mathematics and Physics, University of Surrey, Guildford, Surrey GU2 7XH, United Kingdom}
\author[1]{Praveen C. Srivastava}
\author[3,4,5]{Toshio Suzuki}
	\affil[3]{Department of Physics, College of Humanities and Sciences, Nihon University, Sakurajosui 3, Setagaya-ku,  Tokyo 156-8550, Japan}
	\affil[4]{NAT Research Center, NAT Corporation, 3129-45 Hibara Muramatsu, Tokai-mura, Naka-gun, Ibaraki 319-1112, Japan}
	\affil[5]{School of Physics, Beihang University 100083, People's Republic of China}
 \author[6] {Noritaka Shimizu}
	\affil[6]{Center for Computational Sciences, University of Tsukuba, 1-1-1, Tennodai Tsukuba,\\ Ibaraki 305-8577, Japan}

\date{\hfill \today}

\begin{abstract}
Starting from an isospin-symmetric microscopic interaction of Shin et al. [Phys. Rev. C 110, 034306 (2024)], we developed two isospin symmetry breaking (ISB) shell-model interactions by adding Coulomb interaction, as well as charge-symmetry breaking (CSB) and charge-independence breaking (CIB) components from the Argonne v18 (Av18) interaction at two different radial cutoffs, $r_c = 0.55$ and $0.70\ \mathrm{fm}$. With additional modifications to the proton single-particle energies, we then employed these interactions to calculate the  Coulomb displacement energies (CDEs) and triplet displacement energies (TDEs) of the isobaric multiplet masses for nuclei with $|T_z| = 1/2$ and 
$T = 1$ across the $sd$-shell to test their validity and applicability, and compared with the USDCm interaction. We found  that both 
ISB interactions show important roles of the CSB and CIB components in the CDEs and TDEs.
In addition, we investigated mirror energy differences (MEDs) in the low-energy spectra of nuclei around $A$ = 20, 
and triplet energy differences (TEDs) in the $A$ = 18 and 22 systems. 
They are generally well described by the ISB interactions, although some discrepancies remain.
Furthermore, the calculated $E2$ transition strengths obtained with the
ISB interactions are shown to be in good agreement with experimental data, indicating small ISB effects in these transitions. 
Overall, both the ISB interactions
provide a consistent and physically motivated framework for describing ISB effects, while the  phenomenological USDCm interaction achieves closer agreement with experiment due to a larger
number of adjustable parameters.
\end{abstract}


\maketitle
\section{Introduction}
\label{sect 1}

Atomic nuclei are self-bound quantum many-body systems consisting of protons and neutrons collectively known as nucleons. Due to the similar masses and intrinsic spins of protons and neutrons, they were considered {\color{black}isospin doublets with} different isospin projections \cite{Heisenberg, Wigner}. In nuclear physics, the isospin-symmetry approximation provides an efficient way of developing nucleon-nucleon interactions and simplifying the solution of nuclear many-body problems.  
{\color{black}In addition,}
the nuclear forces 
{\color{black}between} the protons ($T_z$ = -1/2) and the neutrons ($T_z$ = 1/2) are the same within the isospin symmetry approximation. 
For such isospin symmetric nucleon-nucleon interactions, the energy spectra of mirror nuclei, whose protons and neutrons are exchanged, have degenerate excitation spectra. However, isospin symmetry is broken at the nuclear level 
mainly 
{\color{black}because of the} Coulomb interaction between protons, the mass difference between the proton and the neutron, and the charge-dependent nature of {\color{black}the} nuclear force. 
In addition, the experimental data, that is, the difference in the scattering length,  showed that $V_{nn}$ is $(0.6\pm0.4)\%$ more attractive than $V_{pp}$, resulting in charge-symmetry breaking (CSB), and $V_{pn}$ is about 2$\%$ more attractive than ($V_{nn}$ + $V_{pp}$)/2 \cite{brown_89}.
All these effects result in isospin symmetry breaking (ISB) at the nuclear level, and the origin of ISB in nuclear systems can be studied using different nuclear models such as the shell model \cite{brown_85, brown_89, ISB_rev, ISB_sakshi} and the density-functional theory \cite{DFT, ISB_rev2}. This work uses the shell model approach to study ISB effects in the $sd$-shell nuclei. 

The ISB effects provide a way to test different nuclear models and can also help in understanding the nature of nucleon-nucleon forces. These effects are reflected in the isospin-mixing in nuclear states, correction factors needed in the study of $\beta$-decay \cite{ISB_beta1, ISB_beta2}, and in some isospin-forbidden processes \cite{ISB_effect1, ISB_effect2, ISB_effect3}. The mirror energy difference (MED) and {\color{black}the} triplet energy difference (TED) are two key quantities to understand ISB in low energy spectra of atomic nuclei, and {\color{black}both of} these quantities are sensitive to the Coulomb and charge-dependent part of an isospin-symmetry-breaking Hamiltonian. For such ISB Hamiltonians, the degeneracy between the spectra of mirror pairs of nuclei is broken, resulting in a nonzero MED value. While the MEDs for medium mass nuclei in the $fp$-shell are relatively small (around 0.1 MeV) \cite{ISB_pf}, for the lighter mass nuclei in the $sd$-shell can reach near 1 MeV in some cases \cite{ISB_18Ne}. Usually, large MEDs are observed for mirror pairs whose proton-rich partner is near the proton drip line. {\color{black}In such cases, the 1$s_{1/2}$ proton orbital is loosely bound while it is deeply bound on the neutron-rich side}. This is known as the Thomas-Ehrman (TE) shift \cite{TE_1, TE_2}. Accurate modeling of Coulomb and charge-dependent nuclear forces is essential to explain the observed MEDs in mirror pairs of nuclei \cite{ISB_rev3}.
The MED and structure of loosely bound proton-rich $sd$-shell nuclei around $A$ = 20 are discussed in Ref. \cite{MED_A20}. The authors explained the need to reduce some of the two body matrix elements (TBME) to accommodate the weakly bound or unbound effect in such nuclei. The ISB effects within $sd$-shell nuclei and beyond are elaborated in Ref. \cite{isb_sd_2013} by taking into account short-range correlations to the Coulomb interaction and effective charge-dependent force of nuclear origin. Recently, two ISB interactions of USD-type \cite{usd1, usd2, usdb} are available, namely USDC and USDI \cite{usdc,brown_correction}, to study ISB effects within the nuclear shell model.  The ISB interaction USDC is utilized in Ref. \cite{subrajit_M1} to study the contributions of isoscalar and isovector components to the $M1$ transitions of odd-A nuclei in the $sd$-shell. In Ref. \cite{ISB_abinitio}, the \textit{b} and \textit{c} parameters of the isobaric multiplet mass equation (IMME) are calculated using valence space in-medium similarity renormalization group (VS-IMSRG) for $T$ = 1 nuclei with $A$ = 10-74 and  
{\color{black}an overall agreement with the experimental data within approximately 250 keV is obtained}.  In Refs. \cite{MED_sd1, MED_sd2}, the MEDs in the low energy spectra of $sd$-shell nuclei are discussed using VS-IMSRG formalism. Also, in Refs. \cite{ref:Henderson_21, stroberg_2022}, the same VS-IMSRG technique is used to study $E2$ transition strengths within the $sd$-shell nuclei, and it was observed that the $B(E2)$ values for VS-IMSRG show a consistent underprediction compared to the experimental data.  The VS-IMSRG technique and shell model calculations using ISB USD-family interactions have recently been applied to different nuclei in the $sd$-shell to study ISB effects \cite{38Ca_2023, 28S_2022, 22Al_2024, 22Si_2023}.

Recently, in Ref. \cite{Na_NCSM}, we calculated the MEDs in the low-lying states of $|T_z|$ = 1/2 mirror pair ($^{21}$Na - $^{21}$Ne) within the no-core shell model (NCSM) formalism. However, due to computational limitations, calculations were possible only up to $N_{max}$ = 4 model space. In such a scenario, microscopic effective interactions based on NCSM solutions of lower-mass nuclei can be useful \cite{double_ols, sd_int1}.  
Furthermore, ISB versions of 
these microscopic interactions would be beneficial 
to study ISB effects across $sd$-shell nuclei, and they can serve as a complementary set of interactions to newly developed USD-family interactions \cite{usdc}. In this work, we have constructed ISB versions for microscopic interaction, DJ16A (monopole-modified version of DJ16 interaction) \cite{dj16_6A}, to study the ISB effect in low-energy spectra of $|T_z|$ = 1/2 mirror pairs and $T$ = 1 triplets within the $sd$-shell.

This paper is organized as follows. In \autoref{sect 2}, we briefly describe NCSM with a core approach to developing microscopic effective $sd$-shell interaction along with the  Coulomb interaction considered. The \autoref{sect 3} is devoted to presenting the calculated results on IMME, low-energy spectra, and MEDs along with the $E2$ transition strengths. Finally, we summarize our work in \autoref{sect 4}. 

\section{Formalism}
\label{sect 2}
The nuclear shell model stands as a cornerstone of our understanding of nuclear structure,  
providing profound insights into various low-energy properties of atomic nuclei. Central to its framework is the employment of extensive diagonalization within a many-body harmonic oscillator basis. Given the rotational invariance of the nuclear force, it is advantageous to employ single-particle harmonic oscillator states characterized by well-defined quantum numbers: principal quantum number ($n$), azimuthal quantum number ($l$), total angular momentum ($j$), and its z-component ($j_z$). In the traditional shell model, the effective interactions among the valence nucleons are considered in a truncated model space instead of all nucleons being active. Those effective interactions are expressed in terms of single-particle energies (SPE) and TBMEs for different truncated spaces, which are key ingredients of the shell model calculations.

In the early phase of the nuclear shell model, effective interactions with perturbative approaches were developed \cite{Kuo65, Kuo68}. Later, phenomenological interactions for different model spaces have become widely successful in reproducing low-energy spectroscopic data of nuclei in different mass regions \cite{ckpot, usd1, usd2, usdb, kb3g, gxpf1a, Otsuka1,Caurier}. In the last two decades, the \textit{ab initio} methods exhibited great progress and 
these methods provide a way to use chiral interactions from effective field theory directly to study nuclear properties \cite{ncsm, vsimsrg1, vsimsrg2, cc}.  
In addition, using similarity transformations, the \textit{ab initio} methods opened a new path to develop effective interactions for a particular valence space. Two such \textit{ab initio} methods are the NCSM \cite{ncsm,  priyanka1, priyanka2, arch2, chandan} and VS-IMSRG \cite{vsimsrg1, vsimsrg2} methods. Our present work
aims to utilize the effective interactions developed using the former \textit{ab initio} method. 

In Ref. \cite{double_ols}, a new non-perturbative way of developing effective interactions was proposed starting from \textit{ab initio} NCSM wave functions of a few selected lower \textit{p}-shell nuclei. The same technique was recently used to develop effective microscopic interactions for \textit{sd}-shell, namely JISP16 \cite{sd_int1}, N3LO \cite{sd_int1}, and DJ16 \cite{dj16_4A}. These effective interactions were derived from the original interactions \cite{jisp16, n3lo, daejeon16} using the $ab$ $initio$  NCSM wave function and the Okubu-Lee-Suzuki (OLS) \cite{ols_1994} technique. Some of these microscopic interactions have recently been applied to $sd$ shell nuclei in Refs. \cite{priyanka_EPJA, priyanka_NPA, subhrajit_NPA, chandan_PRC} in order to study various different structure properties.

Considering only the $NN$ interaction, the NCSM Hamiltonian can be written as

\begin{eqnarray}
	\label{eq:(1)}
	H_{A}= T_{rel} + V = \frac{1}{A} \sum_{i< j}^{A} \frac{({\vec p_i - \vec p_j})^2}{2M}
	+  \sum_{i<j}^A V^{NN}_{ij}, 
	\end{eqnarray}
where $T_{rel}$ is the relative kinetic energy, $V^{NN}_{ij}$ is the $NN$ interaction containing the nuclear part as well as the Coulomb part, {\color{black}and $M$ is the mass of a nucleon, where the same mass for both proton and neutron is considered.} \autoref{eq:(1)} is in a relative coordinate system, and 
we artificially add the the center of mass (CoM) motion in the Harmonic-oscillator potential ($H_{CM}$) to the original Hamiltonian to restrict the CoM wave function to the $0s$ wave function as

\begin{eqnarray}
	\label{eq:(2)}
	H^{\Omega}_A = H_A + H_{c.m.} = \sum_{i = 1}^A [\frac{p_i^2}{2M} + \frac{1}{2}M \Omega^2 r_i^2]
	+ \sum_{i < j}^A [V^{NN}_{ij} - \frac{M \Omega^2}{2 A} (r_i - r_j)^2].
    \end{eqnarray}

The model space is dependent on a given Harmonic Oscillator (HO) energy $\hbar \Omega$, and a cutoff in the total number of the HO excitation quanta, $N_{max}$. Then NCSM calculations were performed for $^{18}$F (at $N_{max}$ = 6, $\hbar \Omega$ = 14 MeV) using soft $NN$ interaction \cite{daejeon16}, and 28 eigenstates dominated by $N$ = 0 components were considered in constructing an OLS transformation to the $sd$-model space. Furthermore, the NCSM calculations were performed for $^{16}$O to obtain the core energy and for $^{17}$O and $^{17}$F to obtain neutron and proton SPEs. Finally, subtracting the core plus SPEs from the effective Hamiltonian for $^{18}$F yielded TBMEs for the effective $sd$-shell interaction, reported as DJ16$_6$ in Ref. \cite{dj16_6A}. The earlier version of this effective interaction, which is now called DJ16$_4$ \cite{dj16_4A}, was obtained by using OLS two times in contrast to using OLS only once for the newer version. Both the DJ16 interactions given were obtained from the NCSM calculations considering charge symmetry and charge independence of the nuclear force. However, a full charge-dependent version for this effective interaction is possible by deriving effective interactions for $^{18}$O, $^{18}$Ne, and $^{18}$F, separately for the $V_{nn}$, $V_{pp}$ and, $V_{pn}$ parts, respectively \cite{sd_int3} (we have defined these terms more specifically later). 

This work follows a different route to construct ISB versions of the microscopic $sd$-shell effective interaction corresponding to monopole-modified DJ16$_6$ interaction (DJ16$_6$A) from Ref. \cite{dj16_6A}. For simplifying the notation, we call this interaction DJ16A in the rest of the paper. 
The agreement between calculated and experimental energy spectra of about 250 states has been improved for DJ16$_6$ compared to DJ16$_4$; the root-mean-square deviation is reduced to about 600 keV for DJ16A \cite{dj16_6A}.
Starting with the isospin symmetric interaction DJ16A, we intended to obtain suitable ISB versions of this interaction having the form: 
\begin{equation}
    H = H_0 + H_C + H_{INC} ,
\end{equation}
where $H_0$ is the isospin-conserving interaction, and for our case, it is the DJ16A interaction. $H_C$ and $H_{INC}$ are Coulomb interaction and isospin non-conserving part of the nuclear force, respectively. This work employed the Coulomb-CD interaction from Ref. \cite{usdc}, and the TMBEs for this Coulomb interaction were calculated using a harmonic oscillator basis, where the TBMEs depend on mass as $\sim$ $A^{-1/6}$. On the other hand, the TBMEs of the DJ16A interactions are also mass-dependent as $\sim$ $A^{-0.3}$.  All the results presented in this work are calculated within the $sd$-shell model space. It is to be noted OLS transformation is applied only once to get the isospin-symmetric DJ16 interaction. Once we add the Coulomb and isospin non-conserving part of the nuclear force to it, we didn't perform another OLS transformation.

It is well known that the Coulomb interaction is not the only factor responsible for energy splitting in an isobaric multiplet, and it is known as the Nolen-Schiffer anomaly {\color{black} \cite{NS, Suzuki92}}. The charge-dependent nuclear force needs to be considered along with the Coulomb interaction to explain such ISB effects in atomic nuclei. Using USDC and USDCm, 
an increase of the $V_{pn} (T = 1)$ part by 2.2\% and 0.8\%, respectively, is mentioned in Ref. \cite{usdc}. However, instead of increasing the $V_{pn} (T = 1)$ part by a certain amount, the CSB and charge-independence breaking (CIB) effects can be taken into account with the inclusion of isospin-non-conserving interactions in addition to the Coulomb interaction.

In this work, our aim is to test the effect of CSB and CIB interactions in the low-energy spectra of $sd$-shell nuclei. Once the CSB and CIB interactions are considered, the isospin symmetric parts of  $V_{pp}$, $V_{nn}$, and $V_{pn} (T=1)$ are modified as: 

\begin{align*}
V_{pp} =  V_0 + V_{CSB}/2 + V_{CIB}/3\\
V_{nn} = V_0 - V_{CSB}/2 + V_{CIB}/3\\
V_{pn} (T = 1) = V_0 - 2 V_{CIB}/3.
\end{align*}
Here, $V_0$ is the isospin symmetric part of the interaction. In addition to modifying TBMEs, the $V_{CSB}$ also modifies proton SPEs compared to neutron SPEs. On the other hand, $V_{CIB}$ does not affect proton SPEs. 
{\color{black}For that, we consider the CSB and CIB interactions from the {\color{black} Argonne v18 (Av18)} interaction Ref. \cite{AV18_int} 
{\color{black} with 
radial cutoffs ($r_c$): (i) $r_c$ = 0.55 fm and (ii) $r_c$ = 0.70 fm for the short-range components. Thus, the CSB and CIB TBMEs depend on the radial cutoffs.}
{\color{black}
They have the following components with three parameters, $I_{ST}, P_{ST}$, and $Q_{ST}$.
\begin{eqnarray}
V_{ST}(r) &=& I_{ST}T_{\mu}^{2}(r) +[P_{ST} + \mu rQ_{ST}] W(r) \theta(r-r_c), \nonumber\\
T_{\mu}(r) &=& (1+\frac{3}{\mu r} +\frac{3}{(\mu r)^2})\frac{e^{-\mu r}}{\mu r} (1-e^{-cr^2})^2, \nonumber\\
W(r) &=& [1+e^{(r-r_0)/a}]^{-1},
\end{eqnarray}
where $\mu$ = $\frac{1}{3} (m_{\pi^0} + 2m_{\pi^{\pm}} )$ is the average of the pion masses in natural units, $c$ = 2.1 fm$^{-2}$, $r_0$ = 0.5 fm and $a$ = 0.2 fm. 
The $T_{\mu}^2(r)$ term in the first line corresponds to the two-pion exchange force.
A radial cut-off factor, $\theta(r-r_c)$ = 1 (0) for $r\ge r_c$ ($r <  r_c$), is introduced to take into account the short-range correlation effects in the nuclear medium. As the two-pion-exchange term already has a short-range correlation factor, it does not have an extra cut-off factor, which also aplies to the one-pion-exchange term in Eq. (5). The values of the parameters are $I_{S=0, T=1}$ = 0.0 (-0.6024), $P_{S=0, T=1}$ = 3.921 (218.1727), $Q_{S=0, T=1}$ = 2.126 (112.0699), and $I_{S=1, T=1}$ = 0.0 (0.0), $P_{S=1, T=1}$ = 3.921 (0.0), $Q_{S=1, T=1}$ = 2.126 (2.075) for the CSB (CIB) interaction, deduced from TABLE II of Ref. \cite{AV18_int}. 
The CIB interaction also includes one-pion-exchange terms induced by the difference between charged and neutral pion masses. They are expressed as
\begin{eqnarray}
V^{\pi} &=& 2f^2 (v_{\pi}(m_{\pi_0}) -v_{\pi} (m_{\pi_{\pm}})), \nonumber\\
v_{\pi}(m) &=& (\frac{m}{m_{\pi_{\pm}}})^2 \frac{1}{3}mc^2 [Y_{\mu}(r)\sigma_1\cdot\sigma_2 +T_{\mu}(r)S_{12}], \nonumber\\
Y_{\mu}(r) &=& \frac{e^{-\mu r}}{\mu r} (1-e^{-cr^2}),
\end{eqnarray}
where $S_{12}$ = $3(\sigma_1\cdot\hat{\vec{r}})(\sigma_2\cdot\hat{\vec{r}}) -\sigma_1\cdot\sigma_2$, $\mu$ = $m$ in natural units, and $f^2$ = 0.075.
The matrix elements of the CSB and CIB interactions are obtained by using harmonic oscillator wave functions with $\hbar\Omega$ = 14 MeV.}

 The Coulomb displacement energies (CDEs), $\delta\epsilon$, due to the CSB interaction of Av18 are evaluated by $\delta\epsilon_j$ = $\frac{1}{2j+1}\sum_{j_c, J}$ (2J+1)$\langle j j_c:J | V_{CSB} | j j_c: J\rangle$, where $j$ and$ j_c$ are valence and core orbitals, and $J$ = $|j-j_c|$ to $(j+j_c)$. The values of $\delta \epsilon_j$ for $j$=0$d_{5/2}$, 1$s_{1/2}$ and 0$d_{3/2}$ for the $^{17}$F-$^{17}$O pair are obtained to be 90.1, 117.2 and 90.1 keV, respectively, for $r_c = 0.55 fm$, while, while they are 84.8, 110.7, and 84.8 keV for $r_c$ = 0.70 fm. 
{\color{black} The CDEs of the 0$d_{5/2}$ orbital for the $^{17}$F-$^{17}$O pair obtained by the CSB interactions are close to that of 100 keV obtained by the $\rho$-$\omega$ and $\pi$-$\eta$ meson exchange interactions \cite{Suzuki92}.
The experimental CDE (0$d_{5/2}$) for the $^{17}$F-$^{17}$O pair, 3.543 MeV, is found to be well reproduced with the contributions from the meson-exchange CSB interaction within 10 keV for the Hartree-Fock wave functions \cite{Suzuki92}.}

The addition of Coulomb and CSB interactions to the isospin-conserving interaction shifts the SPE of proton orbitals with respect to the corresponding neutron SPEs, along with a reduction in the magnitude of TBME in most cases. 
{\color{black} On the other hand,} the inclusion of CIB interaction shifts the proton and neutron TBMEs by the same amount. 
{\color{black}We consider three sets of interactions for this work based on the Coulomb-CD interaction: \\
(i) DJ16A + CD combines isospin symmetric DJ16A interaction with a Coulomb part without considering the CSB and CIB effects, \\
(ii) DJ16A + CD + Av18 ($r_c$ = 0.55 fm) combine CSB and CIB interactions from Av18 interaction at $r_c$ = 0.55 fm with DJ16A + CD, and \\
(iii) DJ16A + CD + Av18 ($r_c$ = 0.70 fm) is the same as the above interaction (ii), but the CSB and CIB interactions are taken from the Av18 interaction at $r_c$ = 0.7 fm.}
{\color{black} In our model, 
the radial cutoff, which takes into account the short-range correlation effects in the nuclear medium \cite{AB1977, BBW1977, Atti2015} is 
{\color{black} one among very few parameters} to determine the CSB and CIB TBMEs, in contrast to other models, such as in Ref. \cite{Kaneko2017}, where the values of some TBMEs are tuned to reproduce the experimental MED and TED data.
In the case of USDCm \cite{usdc}, the Coulomb TBMEs and SPEs are also tuned to fit the experimental MED data.}

{\color{black}The proton and neutron SPEs for 
the above three interactions are shown in \autoref{tab_spe}, where the same SPEs corresponding to DJ16A + CD + Av18 ($r_c$ = 0.55 fm) and DJ16A + CD + Av18 ($r_c$ = 0.70 fm) are considered.} 
The proton and neutron SPEs for the USDCm interaction are also included in the table.
For the DJ16A+CD+Av18 interactions, the CDEs of 0$d_{5/2}$ and 0$d_{3/2}$ orbitals are taken to be the experimental values, 3.543 and 3.458 MeV, respectively. The experimental CDE for the 1$s_{1/2}$ orbital is 3.167 MeV, which is smaller than that for the 0$d_{5/2}$ orbital by $\sim$400 keV.
The difference in the excitation energies of the 1/2$^+_1$ levels in $^{17}$O and $^{17}$F,  $\sim$400 keV, comes from the Thomas-Ehrman effect, that is, the loose binding of the proton in $^{17}$F \cite{bohr}. The proton separation energy of $^{17}$F is as small as 0.600 MeV.  
 Therefore, instead of adopting the experimental CDE for the 1$s_{1/2}$ orbital, we take the same CDE of 3.543 MeV for the 5/2$^+_1$ and 1/2$^+_1$ states of the ($^{17}$O-$^{17}$F) system.
The proton SPEs are obtained by adding these CDEs to the neutron SPEs of DJ16A \cite{dj16_6A}.
We consider the same proton SPEs for both the second
and the third set of interactions. 
The proton SPEs for the DJ16A + CD are obtained by subtracting the  CDEs of the CSB interaction
from the proton SPEs of  DJ16A + CD + Av18 ($r_c$ = 0.55 fm). 
It is to be noted that the Coulomb, CSB, and CIB SPEs are dependent on $A$ through the single-particle functions.
However, we fix the proton SPEs based on experimental data of the $A$ = 17 pair as explained above.
 The proton SPEs for the USDCm, which are the sum of the neutron SPEs for USDCm and the Coulomb-Fitted SPEs, are taken from Ref. \cite{usdc}.

In order to understand the impact of the Coulomb and (CSB + CIB) TBMEs in modifying the TBMEs of the original isospin-symmetric Hamiltonians, we tabulated {\color{black} in \autoref{tab_tbme}} a few TBMEs involving only $0d_{5/2}$ and $1s_{1/2}$ orbitals corresponding to DJ16A + CD and DJ16A + CD + Av18 ($r_c$ = 0.55 fm). We also calculated $|\text{V}_\text{{pp}}$/$\text{V}_\text{{nn}}|$ for these TBMEs and compared them with the same set of TBMEs for USDCm interaction. Additionally, the reduction factor suggested for the $V_{pp}$ part of USDB interactions is also shown in the seventh column of the table taken from \cite{MED_A20}.
{\color{black}Similar types of reduction in TBMEs due to ISB effects are also reported for $A$ = 16 systems in Ref. \cite{TES_A16_PLB}. }
The table shows that Coulomb and Coulomb + CSB reduce proton-proton TBMEs by  2-38 $\%$ and 
3-38$\%$, respectively,
as shown in the fifth column.} The full interaction files which we have developed are available at Ref. \cite{git_repo}.
 The TBMEs of the CSB and CIB components from Av18 have a mass dependence of $A^{-0.3}$, the same as that of the DJ16A, {\color{black} while those of the Coulomb interaction have a mass dependence of $A^{-1/6}$}.

\begin{table}
\caption{\label{tab_spe}
The SPEs for proton and neutron orbitals (in MeV) corresponding to different ISB interactions are shown.} 
\centering
\begin{tabular}{ccccc}
\hline
Interaction & 0$d_{5/2}$ &  1$s_{1/2}$ & 0$d_{3/2}$\\
 \hline 
Proton SPEs & & & \\ 
\hline
DJ16A + CD {\color{black} ($r_c=$ 0.55 fm}) & {\color{black} -0.4731} & {\color{black} 0.2176} & \color{black} 5.4794
\\[+1pt]
DJ16A + CD + Av18 {\color{black} ($r_c=$ 0.55 fm and 0.70 fm)} & -0.3830 & 0.3348 & 5.5695 \\[+1pt]
USDCm & -0.3836 & 0.1749 & 5.3529\\
\hline
Neutron SPEs & & & \\ 
\hline
DJ16A + CD  & {\color{black} -3.9257} & {\color{black} -3.2079} & \color{black} 2.1117
\\[+1pt]
DJ16A + CD + Av18 	& {\color{black} -3.9257} & {\color{black} -3.2079} & \color{black} 2.1117 \\[+1pt]
USDCm & -3.8958 & -3.1387 & 1.8887\\
\hline
\end{tabular}
\vspace{-10pt}
\end{table}

\begin{table*}
\caption{\label{tab_tbme} {{\color{black}The $V_{pp}$, $V_{nn}$ and $V_{pn}^{T = 1}$ of 
some TBMEs involving only 0$d_{5/2}$ and 1$s_{1/2}$ orbitals corresponding to DJ16A + CD, and DJ16A + CD + Av18 ($r_c$ = 0.55 fm) are shown along with the $|\text{V}_\text{{pp}}$/$\text{V}_\text{{nn}}|$ for these interactions. The $|\text{V}_\text{{pp}}$/$\text{V}_\text{{nn}}|$ for the USDCm interaction \cite{usdc} and the suggested reduction factor for the $\text{V}_{\text{pp}}$ part of USDB$^*$ is taken from \cite{MED_A20} are also shown.}}}
\begin{tabular}{ccccccc}
\hline
 TBME $\langle ij | V | kl \rangle_{JT}$ & V$_{\text{pp}}$ & V$_{\text{nn}}$ & V$_{\text{pn}}^{\textit{T} = 1}$ & $|\text{V}_\text{{pp}}$/$\text{V}_\text{{nn}}|$ & $|\text{V}_\text{{pp}}$/$\text{V}_\text{{nn}}|$  & $|\text{V}_\text{{pp}}$/$\text{V}_\text{{nn}}|$  \\[+2pt]
 \hline
DJ16A + CD & & &  & & USDCm & USDB$^*$\\
\hline \\[-8pt]
$\langle (1s_{1/2})^2 | V |  (1s_{1/2})^2 \rangle_{01}$	&  {\color{black}-1.503}	& {\color{black}-1.910}	& {\color{black}-1.910} & {\color{black}0.79}	&0.73	&0.68\\[+1pt]

$\langle 1s_{1/2} 0d_{5/2} | V |  1s_{1/2} 0d_{5/2} \rangle_{31}$ &  {\color{black}1.115}	& {\color{black}0.751} &	{\color{black}0.751} & {\color{black}1.49}	&1.58	&0.78\\[+1pt]

$\langle 1s_{1/2} 0d_{5/2} | V |  1s_{1/2} 0d_{5/2} \rangle_{21}$ &  {\color{black}-0.644}	&{\color{black}-1.032} & {\color{black}-1.032}	& {\color{black}0.62}	&0.51	&0.84 \\[+1pt]

$\langle (0d_{5/2})^2 | V |  (1s_{1/2})^2 \rangle_{01}$ &  {\color{black}-1.529}	& {\color{black}-1.564} & {\color{black}-1.564} & {\color{black}0.98}	&0.94	&0.80 \\[+1pt]

$\langle 1s_{1/2} 0d_{5/2} | V | (0d_{5/2})^2 \rangle_{21}$ & {\color{black}-0.520}	& {\color{black}-0.544}	& {\color{black}-0.544} & {\color{black}0.96}	&0.89	&0.87\\
 
\hline
DJ16A + CD + Av18 ($r_c$ = 0.55 fm) &  & & & & USDCm & USDB$^*$\\
\hline
$\langle (1s_{1/2}^2) | V |  (1s_{1/2})^2 \rangle_{01}$	&{\color{black}-1.435}	& {\color{black}-1.869} &	{\color{black}-2.019} & {\color{black}0.77}	& 0.73	&0.68
 \\
 
$\langle 1s_{1/2} 0d_{5/2} | V |  1s_{1/2} 0d_{5/2} \rangle_{31}$  & {\color{black}1.126}	& {\color{black}0.757} & {\color{black}0.735} &	{\color{black}1.49} & 1.58	&0.78
 \\[+1pt]
 
$\langle 1s_{1/2} 0d_{5/2} | V |  1s_{1/2} 0d_{5/2} \rangle_{21}$ & {\color{black}-0.644}	& {\color{black}-1.045} &	{\color{black}-1.018} &	{\color{black}0.62} & 0.51	& 0.84 \\[+1pt]

$\langle (0d_{5/2})^2 | V |  (1s_{1/2})^2 \rangle_{01}$ &  {\color{black}-1.500}	& {\color{black}-1.545} &	{\color{black}-1.613} & {\color{black}0.97}	&0.94	&0.80 \\[+1pt]

$\langle 1s_{1/2} 0d_{5/2} | V | (0d_{5/2})^2 \rangle_{21}$ & {\color{black}-0.508} 	& {\color{black}-0.538} &	{\color{black}-0.562} & {\color{black}0.94}	& 0.89	&0.87 \\

\hline
\end{tabular}
\vspace{-10pt}
\end{table*}

\section{Results and Discussions}
\label{sect 3}
\subsection{ {\bf Isobaric multiplet mass equation:}}  
\label{subsect 3.1}

\begin{figure}
\centering
    \includegraphics[scale = 0.5]{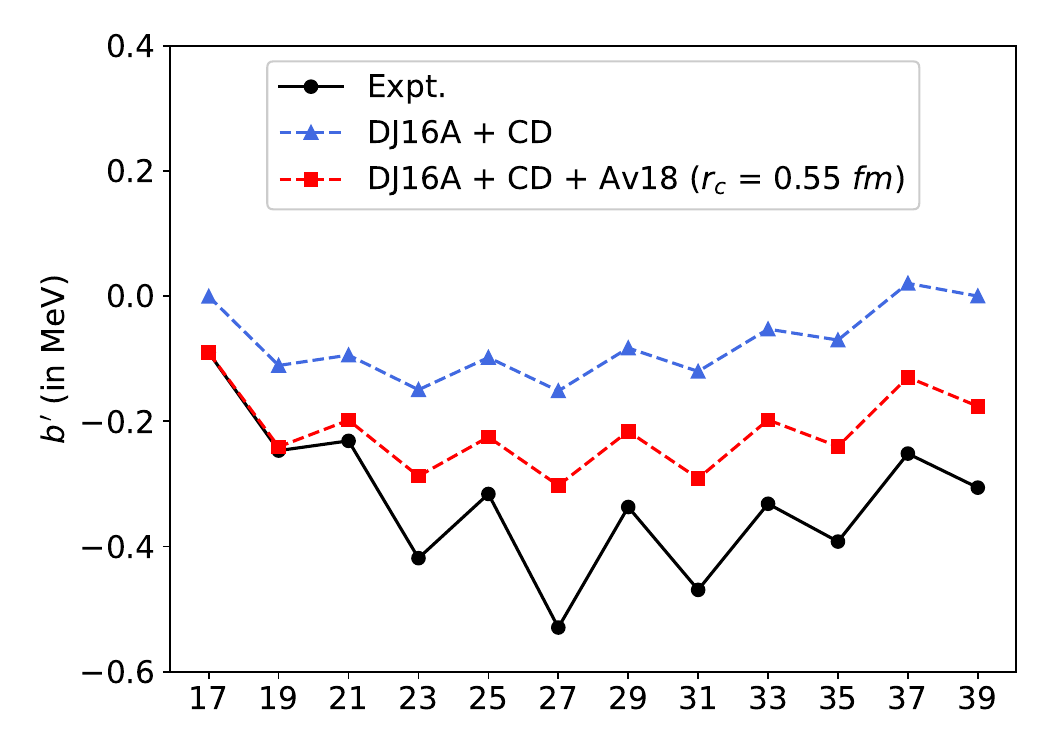}
    \caption{{\color{black}The transformed ($b$) parameters, ($b^\prime$ = $b$ + 3.4526 + $\gamma(A-17)$), are shown for $|T_z|$ = 1/2 mirror pairs. The constant term (3.4526) removes the dominant Coulomb contribution at ($A$ = 17), while the coefficient ($\gamma$) is chosen such that the DJ16A + CD curve becomes zero at ($A$ = 39). The calculated results obtained using two DJ16A-based ISB interactions are compared with the experimental data compiled in Ref.~\cite{AME2020}}.}
    \label{b-fit1}
\end{figure}

\begin{figure}
   \centering
    \includegraphics[scale = 0.48]{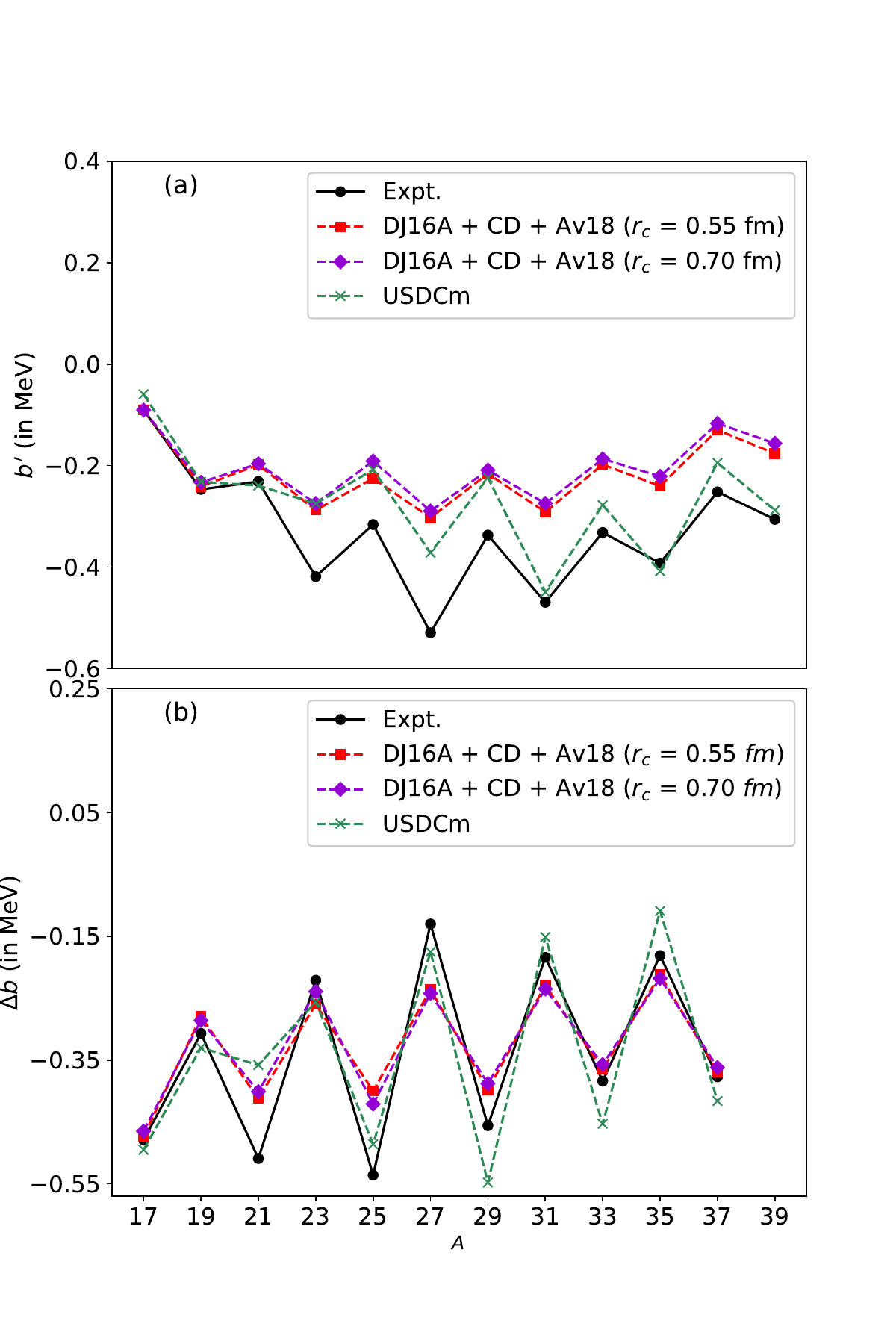}
    \caption{(a) {\color{black}The transformed ($b$) parameters, ($b^\prime$ = $b$ + 3.4526 + $\gamma(A-17)$), are shown for $|T_z|$ = 1/2 mirror pairs. The constant term (3.4526) removes the dominant Coulomb contribution at ($A$ = 17), while the coefficient ($\gamma$) is chosen such that the DJ16A + CD curve becomes zero at ($A$=39). The calculated results obtained using two DJ16A-based ISB interactions are compared with the experimental data compiled in Ref.~\cite{AME2020}}, (b) the calculated $\Delta b$ parameters using the same sets of interactions are compared with the experimental data collected from Ref. \cite{AME2020}. The  $\Delta b$ for $A$ is calculated  as the differences between the $b$ values of $A$ + 2 and $A$ nuclei.} 
    \label{b-fit2}
\end{figure}

The IMME holds the signature of ISB in atomic nuclei. As demonstrated by Wigner \cite{IMME}, the mass difference of isobaric analog states within an isospin multiplet can be expressed as a quadratic form of $T_z$ as:

\begin{equation}
    ME (A, T, T_z) = a + b T_z + c T_z^2,
\end{equation}
where \textit{ME} is mass excess. The $a$, $b$ and $c$ parameters for $|T_z|$ = 1/2 doublets can be written as:
\begin{align*}
a & = [ME (T_z = 1/2) + ME (T_z = -1/2)]/2 \\
b &= ME (T_z = 1/2) - ME (T_z = -1/2)\\
c &= 0.
\end{align*}

The $a$, $b$, and $c$ parameters for $T$ = 1 triplets are written as:
\begin{align*}
a &= ME (T_z = 0) \\
b &= [ME (T_z = 1) - ME (T_z = -1)] / 2 \\
c &= [ME (T_z = 1) + ME (T_z = -1)] / 2 - ME (T_z = 0).
\end{align*}

The $b$- and $c$-parameters correspond to CDEs and triplet displacement energies, respectively.    
In this work, we started with the isospin-symmetric DJ16A interaction from Ref. \cite{dj16_6A}. {\color{black} We then added a Coulomb interaction (Coulomb-CD) from Ref. \cite{usdc} to get the ISB interaction DJ16A + CD. Further, we added the CSB and CIB parts of the Av18 interaction at two radial cutoffs, $r_c$ = 0.55 and 0.70 fm to get two more ISB interactions, namely, DJ16A + CD + Av18 ($r_c$ = 0.55 fm) and DJ16A + CD + Av18 ($r_c$ = 0.70 fm). The proton and neutron SPEs for the first two sets of ISB interactions are shown 
{\color{black} in \autoref{tab_spe}}, and the proton SPEs for DJ16A + CD + Av18 ($r_c$ = 0.70 fm) are considered to be the same as DJ16A + CD + Av18 ($r_c$ = 0.55 fm). 
{\color{black} The experimental $b$-parameter for the 5/2$^+_1$
states of the $A$ = 17 system 
is 3.543 MeV, which is equivalent to the CDE for the $^{17}$F-$^{17}$O mirror pair, is reproduced by the DJ16A+CD+Av18 interactions as their proton SPEs are so chosen.}
}

We now compare the performance of these three sets of ISB interactions in reproducing the experimental $b$-parameters of a few mirror pairs across the $sd$ shell. 
We use the root mean square deviation (rms) of $b$-parameters for that purpose defined as:
	\begin{eqnarray}
	\label{eq:(4)}
	b_{rms} = \sqrt{\frac{1}{N}\sum_{i}^{N}(b_{exp}^i - b_{th}^i)^2}.
	\end{eqnarray} 
 Here, $N$ = 12 for $T_z$ = 1/2 doublets ($A$ = 17-39), and $N$ = 11 for $T_z$ = 1 triplets ($A$ = 18-38). The $b_{rms}$ of the  ground state (g.s.) of $|T_z|$ = 1/2 and $|T_z|$ = 1 nuclei were calculated separately for the DJ16A-based ISB inetractions and 
{\color{black}these} results are compared to the USDCm results.

{\color{black}The DJ16A interaction, combined with Coulomb-CD interaction and CSB and CIB effects from the Av18 interaction, can reproduce the correct ground states of $A$ = 17-39, $|T_z| = 1/2$ doublets, except for $A$ = 25 nuclei. In the case of ($^{25}$Mg - $^{25}$Al), 5/2$^+$ is the g.s. and the calculated results with DJ16A-based  interactions show 1/2$^+$ as the g.s. On the other hand, the USDCm can correctly reproduce 5/2$^+$ as the g.s. of ($^{25}$Mg - $^{25}$Al) pair. Based on the binding energies of the g.s. with respect to the $^{16}$O core, the $b$ parameters are calculated for each pair of $|T_z|$ = 1/2 doublets.
{\color{black} While the $b$-parameter vanishes for the isospin symmetric interaction DJ16A, the modification of the proton SPEs, which is mainly due to the Coulomb interaction between a valence proton and the $^{16}$O-core, changes the $b$-parameter to be $\approx -3.5$ MeV, almost independent of the mass number $A$. The Coulomb interactions among the valence protons in the $sd$ shell lead to a further reduction of the $b$-parameter by an amount proportional to ($A-17$). This amounts to about 3 MeV for the mirror nuclei with large $A$.
Note that the difference in the number of bonds of two-body Coulomb interaction in the valence shells between the nuclei of the mirror pair of mass number $A$ is given by $Z_{1}(Z_{1}-1)/2 -Z_{2}(Z_{2}-1)/2$ =$-\frac{1}{2}(A-17)$, where $Z_{1}=(A-1)/2-8$ and $Z_{2}=(A+1)/2-8$.} 

The experimental and calculated $b$-parameters using DJ16A + CD and DJ16A + CD + Av18 ($r_c$ = 0.55 fm) are compared in \autoref{b-fit1}. Here, the $b$-parameter is obtained by using the calculated g.s. energies except for the ($^{25}$Mg - $^{25}$Al) pair for which the binding energies of 5/2$^+_1$ excited states are considered.
{\color{black} For a better visibility, transformed $b$ parameters, $b^\prime = b + 3.4526 + \gamma(A-17)$, are shown for the $|T_z|$ = 1/2 mirror pairs in \autoref{b-fit1} instead of $b$. 
Here, the constant term, 3.4526, removes the dominant Coulomb contribution at $A$ = 17, while the coefficient $\gamma$ is chosen such that the DJ16A + CD curve becomes zero at $A$ = 39 {\color{black} ($\gamma$ = 0.161)}.} 
From the figure, it can be seen that the Coulomb-CD is not sufficient to reproduce the experimental $b$-parameters. On the other hand, the addition of the CSB and CIB part of the Av18 ($r_c$ = 0.55  fm)  results in better agreement with the experimental data. 

Additionally, we compared the calculated {\color{black}transformed $b$-parameters ($b^\prime$)} using DJ16A + CD + Av18 ($r_c$ = 0.55 fm) and  DJ16A + CD + Av18 ($r_c$ = 0.70 fm) with the USDCm results and the experimental data in the first panel of \autoref{b-fit2}. 
From the figure, it can be seen that the $b^\prime$ parameters, hence the $b$-parameters corresponding to DJ16A + CD + Av18 ($r_c$ = 0.55 fm) and DJ16A + CD + Av18 ($r_c$ = 0.70 fm), are close to each other and comparable to those of USDCm at $A <$30.
The $b$-parameters for the USDCm interaction are closer to the experimental data than for the DJ16A-based interactions at $A\ge$30. This point is also reflected in the $b_{rms}$ values for these three interactions. The $b_{rms}$ for the DJ16A + CD + Av18 ($r_c$ = 0.55 fm), DJ16A + CD + Av18 ($r_c$ = 0.70 fm), and USDCm are 130, 142, and 81 keV, respectively, corresponding to the g.s. of $|T_z| = 1/2$ doublets across $sd$-shell. In the second panel of \autoref{b-fit2}, calculated values of $\Delta b$ ($\Delta b$ for the mass number $A$ is calculated $b$ taking the differences between $b (A + 2)$ and $b (A)$) corresponding to DJ16A + CD + Av18 ($r_c$ = 0.55 fm), DJ16A + CD + Av18 ($r_c$ = 0.70 fm), and USDCm are compared with the experimental data.  The figure shows that DJ16A-based interactions, along with USDCm, can correctly reproduce the experimental staggering pattern of $\Delta b$.

\begin{figure}
   \centering
    \includegraphics[scale = 0.48]{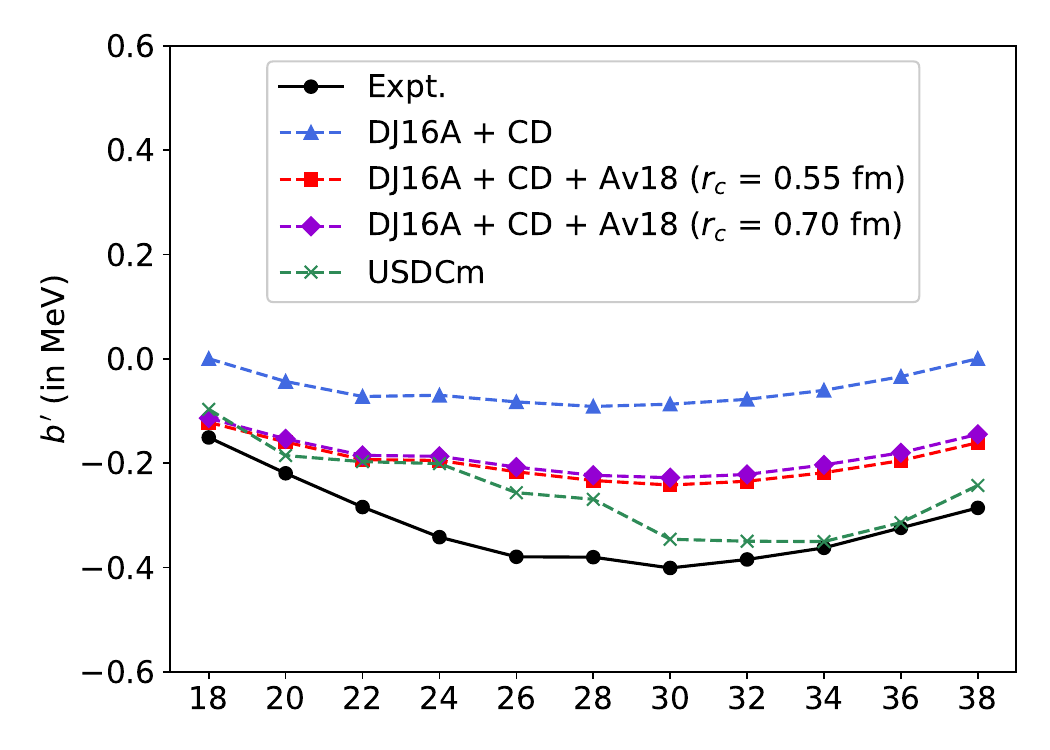}
    \caption{{\color{black}The transformed ($b$) parameters, ($b^\prime$ = $b$ + 3.6820 + $\gamma(A-18)$), are shown for $|T_z|$ = 1 isobaric triplets. The constant term (3.6820) removes the dominant Coulomb contribution at ($A$=18), while the coefficient ($\gamma$) is chosen such that the DJ16A + CD curve becomes zero at ($A$=38). The calculated results obtained using three DJ16A-based ISB interactions are compared with the experimental data compiled in Ref.~\cite{AME2020}}.}
    \label{b-fit3}
\end{figure}

\begin{figure}
  \centering
    \includegraphics[scale = 0.48]{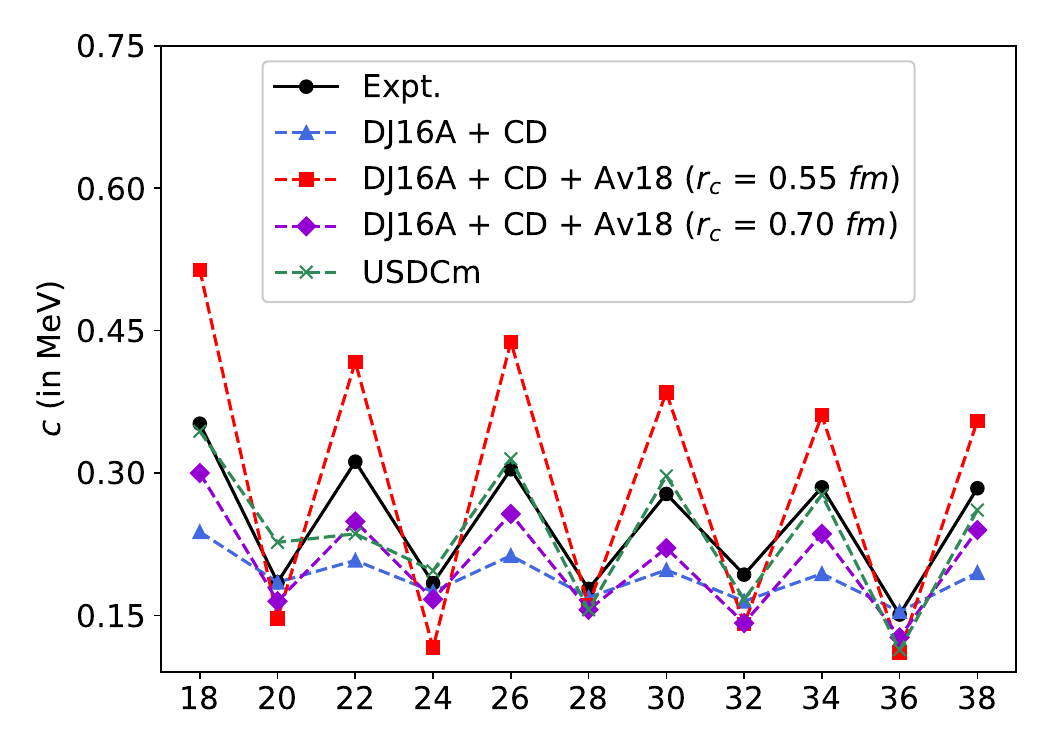}
    \caption{(The calculated $c$ parameters using DJ16A-based ISB interactions, and USDCm are compared with the experimental data collected from Ref. \cite{AME2020} for $T$ = 1 triplet.}
    \label{c-fit1}
\end{figure}

Similarly to the case of $|T_z|$ = 1/2 pairs, the calculated and experimental transformed $b$ parameters ($b^\prime$) for $|T_z|$ = 1 pairs are compared in \autoref{b-fit3}. 
{\color{blue}
{\color{black}  Since $\gamma(A-17)$ = $\gamma$ +$\gamma(A-18)$, the transformed $b$ parameters can be expressed as $ b^\prime$ = $b$ +$b_1$ +$\gamma(A-18)$. Here, we take $b^\prime$ = $b$ + 3.6820 + $\gamma(A-18)$, where the constant term (3.6820) removes the dominant Coulomb contribution at ($A$=18), while the coefficient ($\gamma$) is chosen such that the DJ16A + CD curve becomes zero at ($A$=38)\color{black}($\gamma$ = 0.157)}.}
The DJ16A interaction, combined with Coulomb-CD interaction and CSB and CIB effects from the Av18 interaction, can reproduce the g.s. of $|T_z|$ = 1 pairs from $A$ = 18 to 38 except for $A$ = 24 and 28 pairs. For the case of $A$ = 24 pairs, 4$^+_1$ is the g.s. and calculated results with DJ16A-based interactions showed 1$^+_1$ as the g.s. However, the calculated results with USDCm showed 4$^+_1$ as the g.s. of $A$ = 24 mirror pair consistent with the experimental data. On the other hand, all four ISB interactions, including the USDCm showed 2$^+_1$ as the g.s. of ($^{28}$P - $^{28}$Al) mirror pair instead of the experimental g.s. (3$^+_1$). However, we considered only the experimental g.s. spin parity states while evaluating the $b$-parameters for different ISB interactions. From \autoref{b-fit3}, a similar conclusion can be drawn that Coulomb-CD alone is not sufficient to reproduce the experimental $b$-parameters for $|T_z|$ = 1 pairs. On the other hand, the calculated $b$-parameters using DJ16A + CD + Av18 ($r_c$ = 0.55 fm) and DJ16A + CD + Av18 ($r_c$ = 0.70 fm) are close to each which are slightly away from USDCm results. The $b_{rms}$ values for DJ16A + CD + Av18 ($r_c$ = 0.55 fm) and DJ16A + CD + Av18 ($r_c$ = 0.70 fm) interactions are 129 and 140 keV, respectively, while it is 77 keV for USDCm. Based on the $b_{rms}$ values of the g.s. of $|T_z|$ = 1/2 and $|T_z|$ = 1 nuclei across the $sd$-shell, all three ISB interactions showed $b_{rms}$ within the range 77 to 142 keV up to $A$ = 39 mirror pair.

\begin{figure*}
    \centering
    \includegraphics[scale = 0.45]{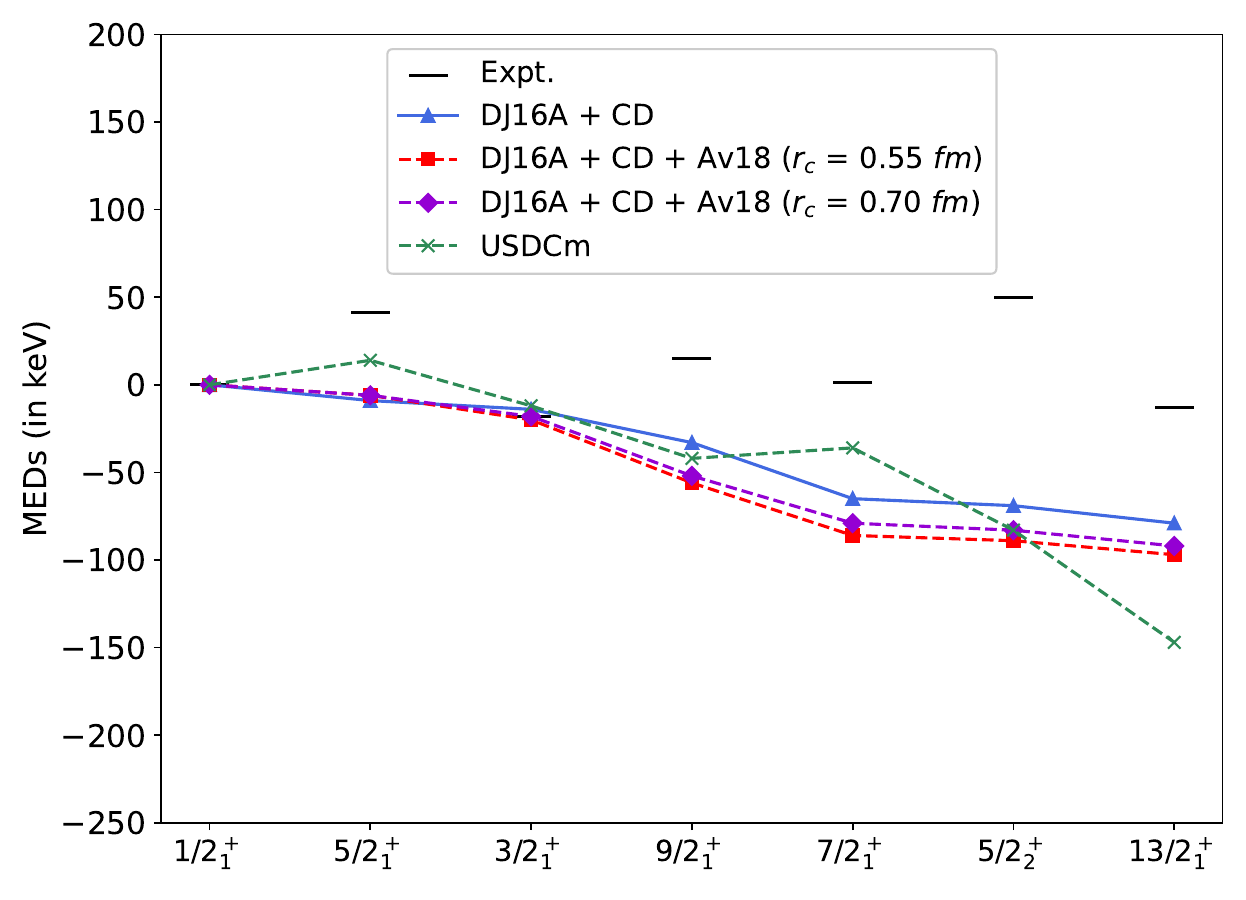}
    \caption{Calculated MEDs for low-lying states of mirror pair ($^{19}$Ne - $^{19}$F), using {\color{black}DJ16A + CD,} DJ16A + CD + Av18 ($r_c$ = 0.55 fm), DJ16A + CD + Av18 ($r_c$ = 0.70 fm) and USDCm interactions are shown.}
    \label{A19_occ}
\end{figure*}

\begin{figure*}
    \centering
    \includegraphics[scale = 0.45]{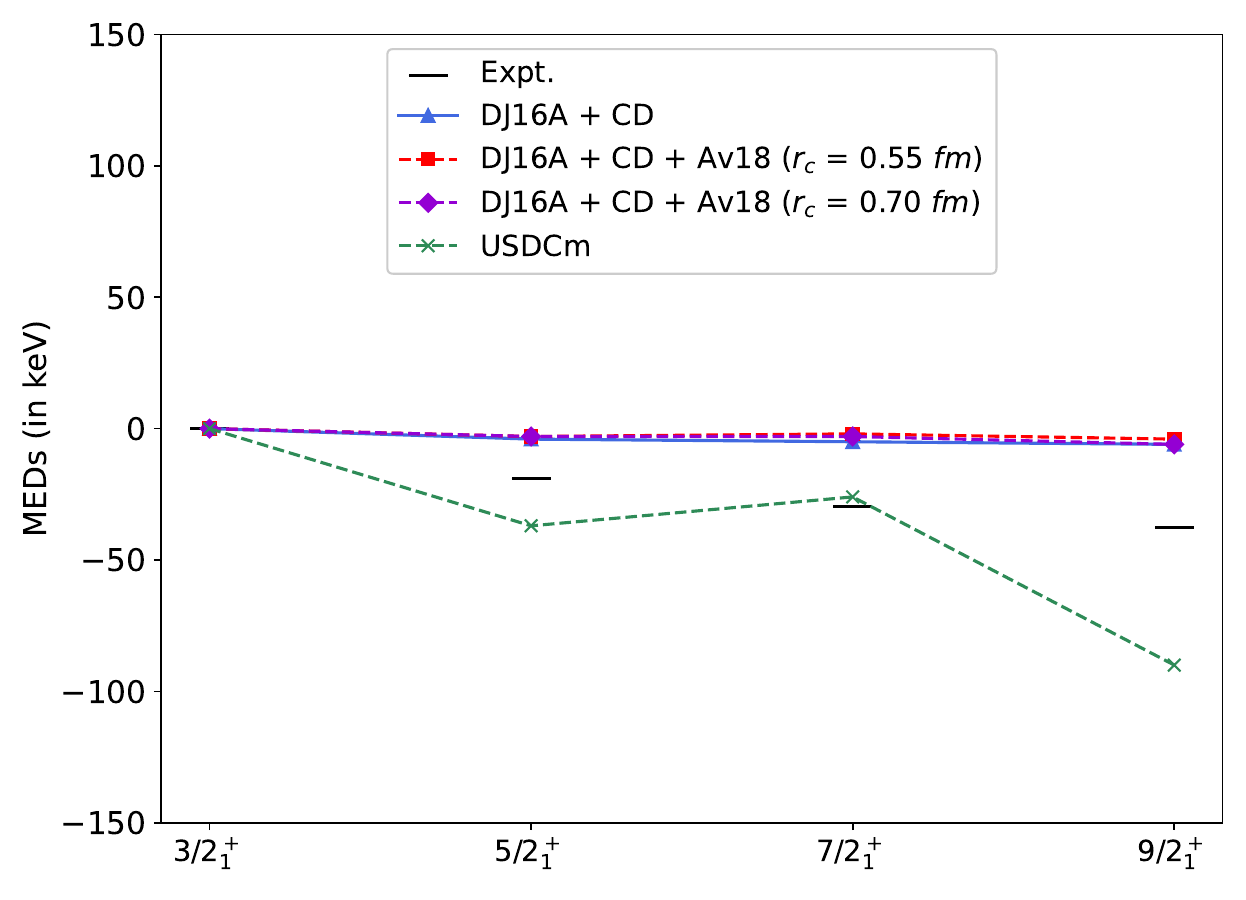}
    \caption{Calculated  MEDs for low-lying states of mirror pair ($^{21}$Na - $^{21}$Ne), using {\color{black}DJ16A + CD,} DJ16A + CD + Av18 ($r_c$ = 0.55 fm), DJ16A + CD + Av18 ($r_c$ = 0.70 fm) and USDCm interactions are shown.}
    \label{A21_occ}
\end{figure*}

Furthermore, we calculate the $c$-parameters for the $T$= 1 nuclei in the $sd$ shell. Firstly, we calculated the $c$-parameters for DJ16A + CD, and those results are significantly less than the experimental $c$-parameters as shown in \autoref{c-fit1}, suggesting the need to include the CIB effects. For example, DJ16A + CD results in $c(A = 18)$ = 237 keV instead of the experimental value of 352 keV. We, therefore, added the CIB contributions from the Av18 interaction. From the figure, it can also be seen that the calculated $c$ parameters can reproduce the staggering pattern of the experimental data for all the ISB interactions considered. The $c_{rms}$ for DJ16A + CD + Av18 ($r_c$ = 0.55 fm), DJ16A + CD + Av18 ($r_c$ = 0.70 fm), and USDCm are 98, 41, and 37 keV, respectively. Between the two DJ16A-based ISB interactions, the DJ16A + CD + Av18 ($r_c$ = 0.55 fm) shows overestimation of the $c$-parameters for the g.s in a few cases.

\subsection{{\bf Mirror energy differences }} 

The effect of isospin symmetry breaking can also be seen in the low-energy spectra of mirror nuclei. The MED provides an estimation of the ISB in low-energy spectra, and it is defined as:

\begin{eqnarray}
	MED_J = E_J(T_z = -T) - E_J(T_z = +T).
	\end{eqnarray}
Here, $E_J$ are the excitation energies of isobaric analog states with angular momentum $J$ in a mirror pair with $T_z$ = $\pm T$.  
{\color{black} Among the eleven $|T_z| = 1/2$ doublets, we selected the ($^{19}$Ne - $^{19}$F) and ($^{21}$Na - $^{21}$Ne) mirror pairs, along with two $T$ = 1 triplets, ($^{18}$Ne/$^{18}$F/$^{18}$O) and ($^{22}$Mg/$^{22}$Na/ $^{22}$Ne). The low-energy spectra of these four sets of nuclei are shown in \autoref{tab:2tz_1_spectra} and \autoref{tab:2tz_2_spectra}, and their corresponding MEDs are shown in Figures \ref{A19_occ}-\ref{A22_occ}.}
{\color{black} The r.m.s. deviations of the calculated energies from the experimental values are also denoted in the Tables.
They are generally smaller for the USDB and USDC than the DJ16A.}
{\color{black}The CSB and CIB interactions are found to reduce the deviations in some cases, particularly for the $A$ = 18 and 22 nuclei.}

\begin{table*}
\caption{\label{spectra_doublets} {
{\color{black}Low-energy spectra of $A$ = 19 and 21 doublets are shown  in unit of MeV.}}
}
\label{tab:2tz_1_spectra}
\begin{tabular}{ccccccccc}
\hline
$J^\pi$ & Expt. &USDB & DJ16A & DJ16A(1) & DJ16A(2) & DJ16A(3)  & DJ16A(4) & USDCm \\
 \hline 
 $^{19}$Ne   &  &  &  &  &  &  & &\\
1/2$^+_1$  & 0 & 0 & 0 & 0 & 0 & 0 & 0 & 0\\

 5/2$^+_1$ & 0.23827(11)& 0.077 & 0.354 & 0.352 & 0.350 & 0.352 & 0.352 & 0.075 \\

 3/2$^+_1$ & 1.5360(4) & 1.770 & 1.908 & 1.901 & 1.881 & 1.879 & 1.879 & 1.758 \\

 9/2$^+_1$ & 2.7947(6) & 2.599 & 2.910 & 2.911 & 2.887 & 2.875 & 2.873 & 2.577\\

 7/2$^+_1$ & 4.3791(22) & 4.868 & 5.181 & 5.190 & 5.126 & 5.128 & 5.119 & 4.897 \\

 5/2$^+_2$ & 4.600(4) & 4.969 & 5.321 & 5.329 & 5.261 & 5.259 & 5.254 & 4.940\\

 13/2$^+_1$ & 4.635(4) & 4.604 & 4.956  & 4.960 & 4.889 & 4.879 & 4.883 & 4.502\\
\hline
 r.m.s.  & & 0.287 & 0.488 & 0.492 & 0.447 & 0.446 & 0.442 & 0.296\\
  deviation & &  &  &  &  &  &  & \\
 
\hline 
$^{19}$F & &  & & & & & &  \\
 1/2$^+_1$ & 0 & 0 & 0 & 0 & 0 & 0 & 0 & 0\\

 5/2$^+_1$ & 0.197143(4) & 0.077 & 0.354 & 0.359 & 0.359 & 0.358 & 0.358 & 0.061\\

 3/2$^+_1$ & 1.554038(9) & 1.770 & 1.908 & 1.895 & 1.895 & 1.899 & 1.897 & 1.770\\

 9/2$^+_1$ & 2.7779849(34) & 2.599 & 2.910 & 2.920 & 2.920 & 2.931 & 2.925 & 2.619\\

 7/2$^+_1$ & 4.377700(42) & 4.868 & 5.181 & 5.191 & 5.191 & 5.214 & 5.198 & 4.933\\

 5/2$^+_2$ & 4.5499(8) & 4.969 & 5.321 & 5.330 & 5.330 & 5.348 & 5.337 & 5.023\\

 13/2$^+_1$ & 4.648(1) & 4.604 & 4.956 & 4.968  & 4.968 & 4.976 & 4.975 & 4.649\\

\hline
 r.m.s. & & 0.292 & 0.500 & 0.506 & 0.506 & 0.518 & 0.511 & 0.323 \\

 deviation & &  &  &  &  &  &  & \\
\hline

$^{21}$Na & &  & & & & & & \\
3/2$^+_1$ & 0 & 0 & 0 & 0 & 0 & 0 & 0 & 0\\

 5/2$^+_1$ & 0.33190(10) & 0.266 & 0.541 & 0.541 & 0.534& 0.531 & 0.534 & 0.254\\

 7/2$^+_1$ & 1.7161(3) & 1.757 & 1.852 & 1.852 & 1.836 & 1.833 & 1.837 & 1.755\\

 9/2$^+_1$ & 2.8291(7) & 2.831 & 3.135 & 3.137 & 3.113 & 3.107 & 3.113 & 2.757 \\

\hline
 r.m.s. & & 0.045 & 0.228 & 0.229 & 0.213 & 0.209 & 0.213 & 0.065\\
 deviation & &  &  &  &  &  &  & \\
\hline

$^{21}$Ne & &  & & & & & & \\
 3/2$^+_1$ & 0 & 0 & 0 & 0 & 0 & 0 & 0 & 0\\

 5/2$^+_1$ & 0.350727(8) & 0.266 & 0.541 & 0.540 & 0.538 & 0.534 & 0.537 & 0.291\\

 7/2$^+_1$ & 1.745910(19) & 1.757 & 1.852 & 1.851 & 1.841& 1.835 & 1.840 & 1.781\\

 9/2$^+_1$ & 2.8666(2) & 2.831 & 3.135 & 3.133 & 3.119 & 3.111 & 3.119 & 2.847\\

\hline
 r.m.s.  & & 0.053 & 0.200 & 0.198 & 0.190 & 0.184 & 0.189 & 0.042 \\
 deviation & &  &  &  &  &  &  & \\

\hline

\end{tabular}
\end{table*}

\begin{table*}
\caption{\label{spectra_triplets} {{\color{black}Low-energy spectra of $A$ = 18 and 22 triplets are shown.}}}
\label{tab:2tz_2_spectra}
\resizebox{16cm}{8cm}{
\begin{tabular}{ccccccccc}
\hline
 $J^\pi$ & Expt. & USDB & DJ16A & DJ16A(1) & DJ16A(2) & DJ16A(3)  & DJ16A(4) & USDCm  \\
 \hline 
$^{18}$Ne &  &  &  &  &  &  & &\\

 0$^+_1$ & 0 & 0 & 0 & 0 & 0 & 0 & 0 & 0 \\

 2$^+_1$ & 1.8873(2) & 1.998 & 2.010 & 2.011 & 1.957 & 1.867 & 1.923 & 1.972\\

 4$^+_1$ & 3.3762(4) & 3.527 & 3.434 & 3.436 & 3.334 &3.229 & 3.297 & 3.430\\

\hline
 r.m.s.  & & 0.132 & 0.096 & 0.097 &  0.058 & 0.195 & 0.061 & 0.071\\
 deviation & &  &  &  &  &  &  & \\

\hline
$^{18}$F &  &  &  &  &  &  & &\\

 0$^+_1$ & 1.04155(8) & 1.481 & 1.700 & 1.527 & 1.527 & 1.527 & 1.671 & 1.283\\

 2$^+_2$ & 3.06184(18) & 3.480 & 3.709 & 3.707 & 3.707 & 3.707 & 3.727 & {\color{black}3.323}\\

 4$^+_1$ & 4.652(2) & 5.009 & 5.134 & 5.128 & 5.128 & 5.128 & 5.149 & 4.963\\

\hline
r.m.s. & & 0.406 & 0.601 & 0.541 & 0.541 &  0.541 & 0.602 & 0.273\\
deviation & &  &  &  &  &  &  & \\

\hline 
$^{18}$O &  &  &  &  &  &  & &\\

 0$^+_1$ & 0 & 0 & 0 & 0 & 0 & 0 & 0 & 0\\

 2$^+_1$ & 1.98207(9) & 1.998 & 2.010 &2.010 & 2.010 & 1.953 & 1.998 & 2.027\\

 4$^+_1$ & 3.55484(40)& 3.527 & 3.434 & 3.434 & 3.434 & 3.374 & 3.428 & 3.611\\

\hline
r.m.s. & & 0.023 & 0.088 &  0.088 & 0.088 &  0.130  &  0.090 & 0.051\\
deviation & &  &  &  &  &  &  & \\

\hline 
$^{22}$Mg &  &  &  &  &  &  && \\

 0$^+_1$ & 0 & 0 & 0 & 0 & 0 & 0 & 0 & 0\\

 2$^+_1$ & 1.24702(3) & 1.363 & 1.129 & 1.130 & 1.115 & 1.092 & 1.109 & 1.331\\

 4$^+_1$ & 3.30822(6) & 3.357 & 3.259 & 3.262 & 3.225 & 3.166 & 3.209 & 3.318\\

 2$^+_2$ & 4.4020(3) & 4.284 & 4.941 & 4.928 & 
4.901 & 4.845 & 4.885 & 4.300\\
\hline
r.m.s. & & 0.100 & 0.320 & 0.312 &  0.301 & 0.283 & 0.296 & 0.077\\
deviation & &  &  &  &  &  &  & \\

\hline
$^{22}$Na &  &  &  &  &  &  & &\\

 0$^+_1$ & 0.65700(14) & 0.865 & 0.872 & 0.727 & 0.727 & 0.727 & 0.843 & 0.784\\

 2$^+_1$ & 1.9518(3) & 2.228 & 2.001 & 1.898 & 1.898 & 1.898 & 1.981 & 2.146\\

 4$^+_2$ & 4.0714(6)& 4.222 & 4.132 & 4.079 & 4.079 & 4.079 & 4.117 & 4.151\\

 2$^+_4$ & 5.0625(4)& 5.148 & 5.814 & 5.751 & 5.751 & 5.751 & 5.797 & 5.108\\

\hline
 r.m.s. & & 0.193 & 0.393 & 0.347 &  0.347 & 0.347 & 0.380 & 0.126\\
 deviation & &  &  &  &  &  &  & \\

\hline 
$^{22}$Ne &  &  &  &  &  &  & &\\

 0$^+_1$ & 0 & 0 & 0 & 0 & 0 & 0 & 0 &0 \\

 2$^+_1$ & 1.274537(7) & 1.363 & 1.129 & 1.128 & 1.120 & 1.100 & 1.116 & 1.340\\

 4$^+_1$ & 3.3572(5) & 3.357 & 3.259 & 3.255 & 3.233 & 3.179 & 3.221 & 3.318\\

 2$^+_2$ & 4.4562(9) & 4.284 & 4.941 & 4.944 & 4.927 & 4.880 & 4.918 & 4.296\\

\hline
 r.m.s. & & 0.112 & 0.300 & 0.299 &  0.295 & 0.284 & 0.293 & 0.102\\
 deviation & &  &  &  &  &  &  & \\
\hline
\end{tabular}
}
\vspace{-10pt}
\end{table*}

\textbf{$^{19}$Ne/$^{19}$F:} {\color{black} The low-energy spectra of ($^{19}$Ne - $^{19}$F) mirror pair is shown in \autoref{tab:2tz_1_spectra}. {\color{black}The shell model calculations were performed using {\color{black}isospin symmetric USDB and DJ16A interactions and compared them with DJ16A-based ISB interactions, namely DJ16A(1), DJ16A(2), DJ16A(3), and DJ16A(4) , and recently developed USDCm interactions. Here, DJ16A(1) is the isospin symmetric TBMEs of DJ16A with modified proton SPEs according to  \autoref{tab_tbme} 
and DJ16A(2) is the DJ16A + CD with modified proton SPEs without CSB and CIB effects. Finally, DJ16A(3) and DJ16A(4) are DJ16A + CD + Av18 ($r_c$ = 0.55 fm) and DJ16A + CD + Av18 ($r_c$ = 0.70 fm) interactions, respectively}. The calculated spectra follow the correct ordering as the experimental data up to the third excited state (9/2$^+_1$). Experimentally, the first excited state (5/2$^+_1$) of $^{19}$Ne and $^{19}$F are at 0.238 and 0.197 MeV, respectively. {\color{black}While the USDCm interaction shows excitation energies of 5/2$^+_1$ state around 70 keV for both nuclei, the DJ16A + CD + Av18 ($r_c$ = 0.55 fm) and DJ16A + CD + Av18 ($r_c$ = 0.70 fm) interactions show excitation energies near to 350 keV.  
{\color{black}The calculated MEDs in the low-energy spectra of ($^{19}$Ne - $^{19}$F) mirror pair corresponding to three DJ16A-based interactions and the USDCm interaction are compared to the experimental data in \autoref{A19_occ}.} 
Up to the third excited state ($9/2^+_1$), the proton and neutron occupancies in the valence shells are nearly identical and primarily dominated in the valence $0d_{5/2}$ orbitals for all three interactions. 
This leads to small MEDs for these states that agree with the experimental data within an error range of 70 keV. On the other hand, the calculated MEDs for 7/2$^+_1$, 5/2$^+_2$, and 13/2$^+_1$  overpredict the experimental data. The MEDs corresponding to both the DJ16A-based interactions agree with each other, with a maximum deviation of only 7 keV.

\textbf{$^{21}$Na/$^{21}$Ne:} The calculated low-energy spectra of ($^{21}$Na - $^{21}$Ne) mirror pair is shown in \autoref{tab:2tz_1_spectra} 
{\color{black}together} with the experimental data. {\color{black} All seven interactions, including the isospin symmetric USDB and DJ16A, employed in this work can reproduce the correct ordering up to the third excited state (9/2$^+_1$). While the first excited state (5/2$^+_1$) obtained for USDCm is slightly less than the experimental data, the DJ16A-based interactions show that this state is 200 keV higher than the experimental data. On the other hand, the second excited state (7/2$^+_1$) obtained for all ISB interactions is in good agreement with the experimental data. }{\color{black}In \autoref{A21_occ}, a comparison between the calculated and experimentally observed MEDs is 
made
for  {\color{black}DJ16A + CD}, DJ16A + CD + Av18 ($r_c$ = 0.55 fm),  DJ16A + CD + Av18 ($r_c$ = 0.70 fm), and USDCm interactions. {\color{black} While the contributions from the CSB part of the interactions are tiny for the present pair,} the calculated MEDs for the 5/2$_{1}^{+}$ and $7/2_{1}^{+}$ are consistent with the experimental values within the error range of $\sim$ 28 keV. The calculated MED of the 9/2$^+_1$ state for the DJ16A-based interactions 
{\color{black} deviates from}
the experimental data 
{\color{black} by} $\sim$ 33 keV, while for the USDCm interaction, it is near 52 keV. }

\begin{figure*}
    \centering
    \includegraphics[scale = 0.45]{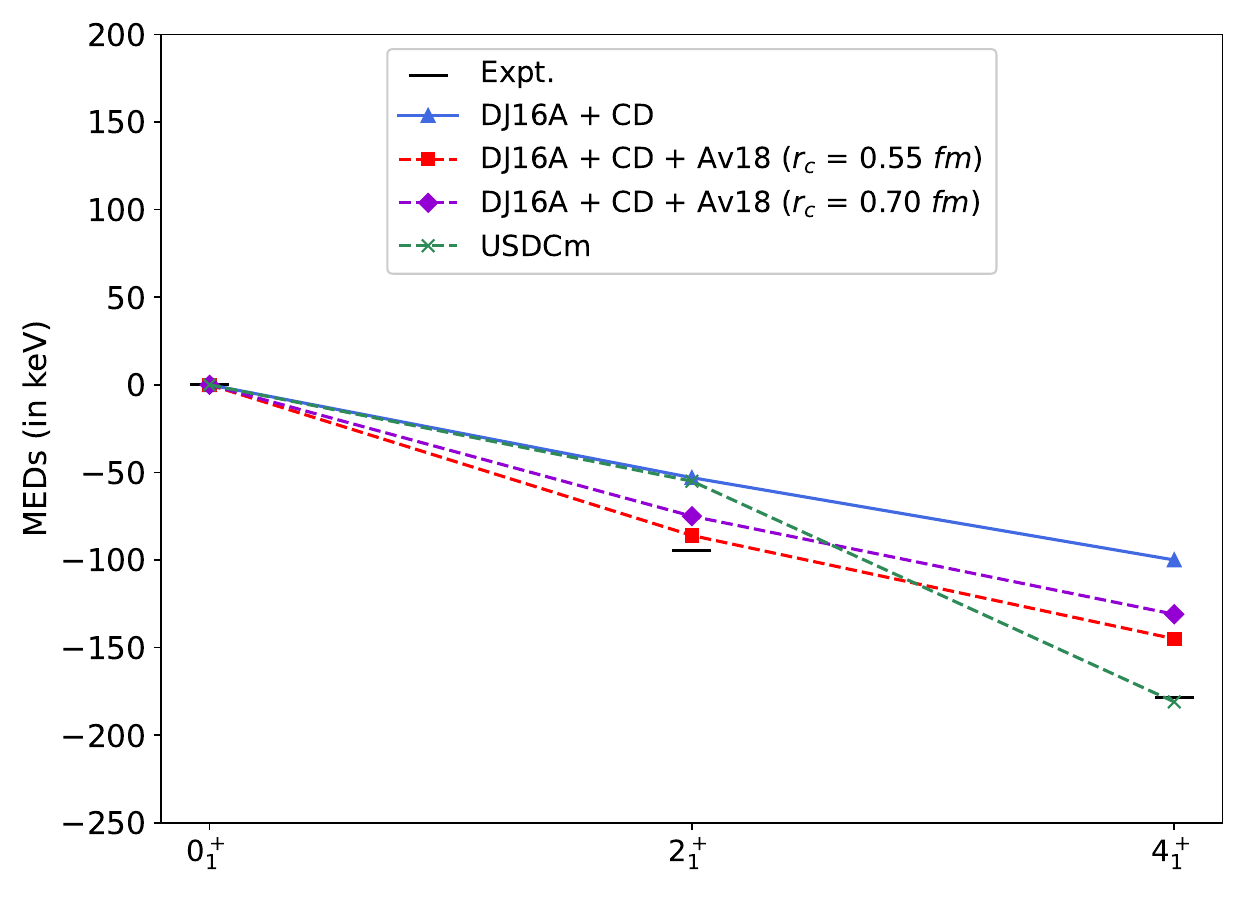}
    \caption{Calculated MEDs for low-lying states of mirror pair ($^{18}$Ne - $^{18}$O), using  {\color{black}DJ16A + CD,} DJ16A + CD + Av18 ($r_c$ = 0.55 fm), DJ16A + CD + Av18 ($r_c$ = 0.70 fm) and USDCm interactions are shown.}
    \label{A18_occ}
\end{figure*}

\begin{figure*}
    \centering
    \includegraphics[scale = 0.45]{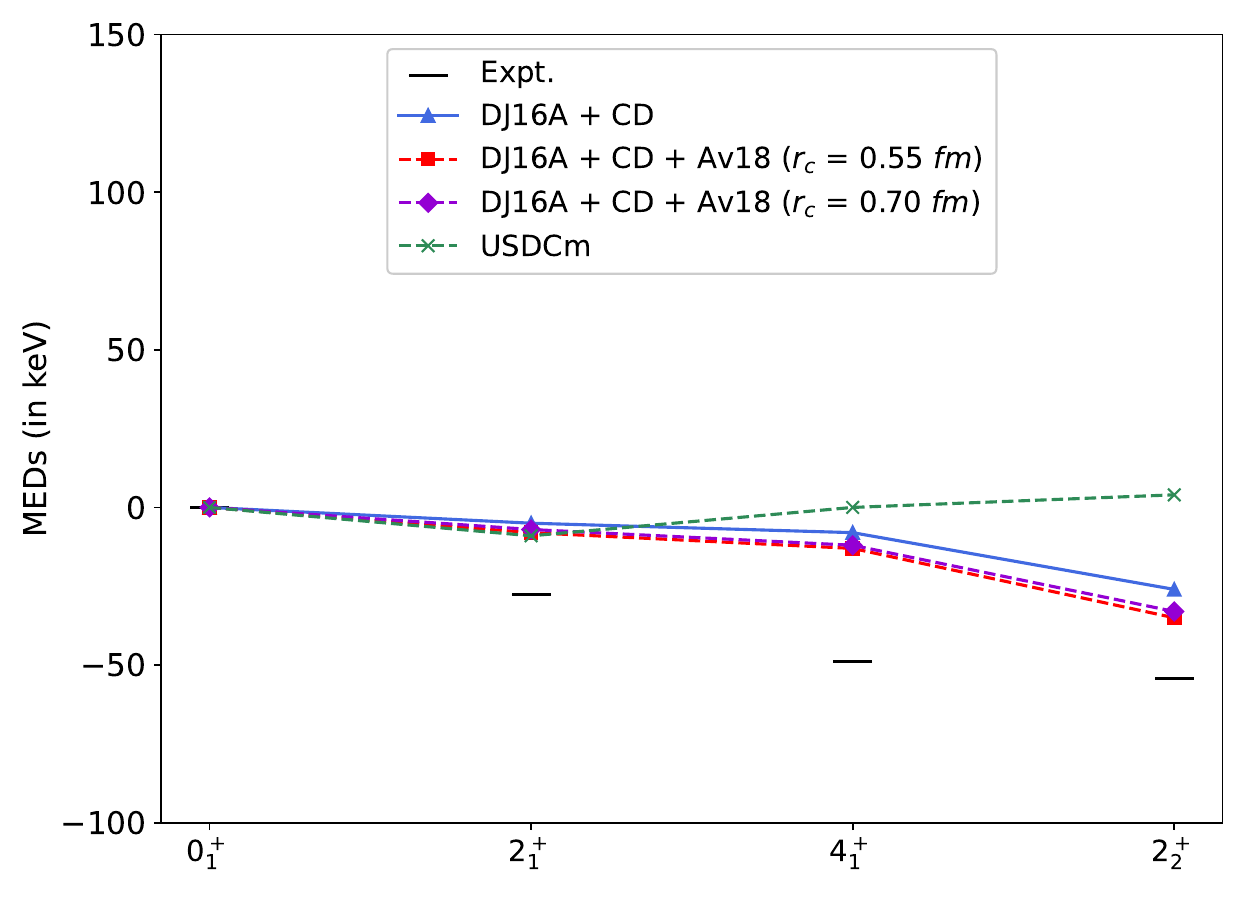}
    \caption{Calculated MEDs for low-lying states of mirror pair ($^{22}$Mg - $^{22}$Ne), using  {\color{black}DJ16A + CD,} DJ16A + CD + Av18 ($r_c$ = 0.55 fm), DJ16A + CD + Av18 ($r_c$ = 0.70 fm) and USDCm interactions are shown.}
    \label{A22_occ}
\end{figure*}
\textbf{$^{18}$Ne/$^{18}$F/$^{18}$O:}  {\color{black}In \autoref{tab:2tz_2_spectra}, the calculated energy spectra comprising 
three low-lying states corresponding to $|T_z|$ = 0, 1  triplet ($^{18}$Ne - $^{18}$F - $^{18}$O) is compared to the experimental data. The excitation energies of the first and second excited states of the $|T_z|$ = 1 mirror pair, obtained for DJ16A-based ISB interactions and USDCm, are in good agreement with the experimental data. On the other hand, the excitation energies of $^{18}$F calculated using DJ16A-based interactions are higher than the experimental data. However, the USDCm results show a good agreement with the experimental data.}  In \autoref{A18_occ}, a comparison between the calculated and experimentally observed MEDs is made.  {\color{black}Calculated MEDs for the $2_{1}^{+}$ and $4_{1}^{+}$ states using  DJ16A + CD + Av18 ($r_c$ = 0.55 fm), DJ16A + CD + Av18 ($r_c$ = 0.70 fm) and USDCm are rather close to the experimental MEDs within the range of $\sim$ 48 keV. }

\textbf{$^{22}$Mg/$^{22}$Na/$^{22}$Ne:} {\color{black}The calculated low-energy spectra of ($^{22}$Mg - $^{22}$Na - $^{22}$Ne) triplet is shown in \autoref{tab:2tz_2_spectra} along with the experimental data. It can be seen that the calculated spectra of $A$ = 22 nuclei using  DJ16A + CD + Av18 ($r_c$ = 0.55 fm), DJ16A + CD + Av18 ($r_c$ = 0.70 fm), and USDCm are in good agreement with the experimental data up to the second excited state. However, the excitation energies of the 2$^+_2$ for ($^{22}$Mg - $^{22}$Ne) and 2$^+_4$ for $^{22}$Na obtained for both the DJ16A-based interactions are almost 500 and 600 keV higher than the experimental excitation energies. On the other hand, USDCm shows good agreement with the experimental data for these states.} In \autoref{A22_occ}, a comparison is shown between the calculated and experimentally measured MEDs. {\color{black}While the calculated MEDs of the three excited states corresponding to the USDCm interaction are less 
than the experimentally observed MEDs, the DJ16A + CD + Av18 ($r_c$ = 0.55 fm)  interaction shows better agreement with the observed data. Overall, the calculated MEDs for the three excited states are fairly close to the experimental MEDs, with differences within 59 keV.}

\begin{figure}
    \centering
    \includegraphics[scale = 0.55]{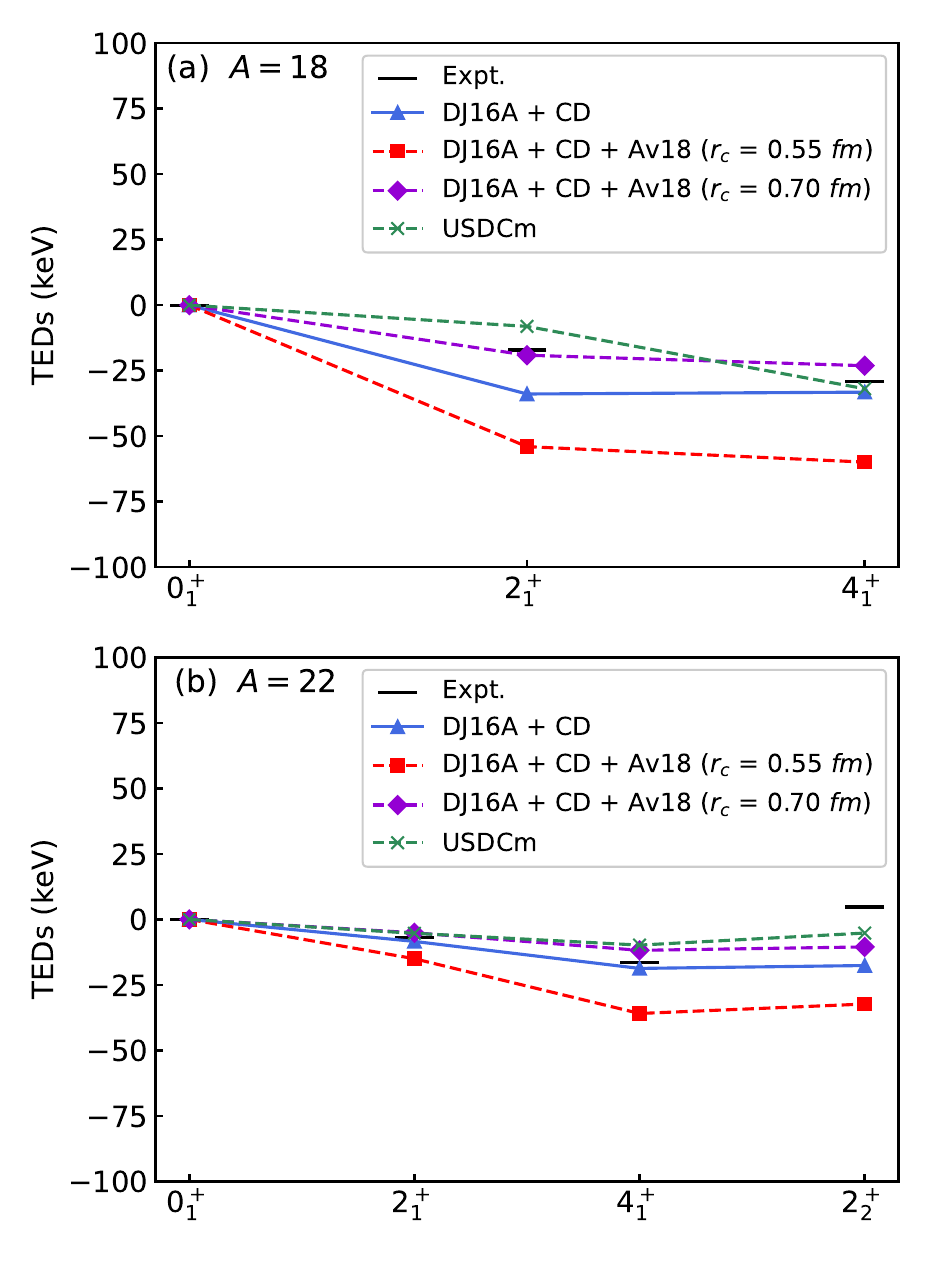}
    \caption{Calculated TEDs for low-lying states   of (a) $A$ = 18 ($^{18}$Ne/$^{18}$F/$^{18}$O), and (b) $A$ = 22 ($^{22}$Mg/$^{22}$Na/$^{22}$Ne) systems using  {\color{black}DJ16A + CD,} DJ16A + CD + Av18 ($r_c$ = 0.55 fm), DJ16A + CD + Av18 ($r_c$ = 0.70 fm) and USDCm interactions are shown.}
    \label{TED}
\end{figure}

\subsection{{\bf Triplet energy differences}}
The TED is another quantity apart from the MEDs that also show the isospin symmetry breaking effects within $T$ = 1 triplets. It is defined as:

\begin{eqnarray}
TED_J &=& [E_J(T_z = -1) + E_J(T_z = +1)]/2 \nonumber  - E_J(T_z = 0).
\end{eqnarray}
Here, $E_J$ are the excitation energies of isobaric analog states with angular momentum $J$ in a $T$ = 1 triplet with $T_z$ = 0, $\pm 1$.  

From the spectra data presented in \autoref{tab:2tz_2_spectra}, we calculated the TEDs for the $A$ = 18 and 22 triplets, and these results are shown in \autoref{TED}. From the figure, it can be seen that for both cases, the results of DJ16A + CD + Av18 ($r_c$ = 0.70 fm) is comparable to the USDCm results and the experimental data. On the other hand,  {\color{black}both DJ16A + CD and} DJ16A + CD + Av18 ($r_c$ = 0.55 fm) show comparatively large deviations for both cases.

\subsection{{\bf Electromagnetic properties of mirror nuclei}}

The $E2$ transition strengths between different states of nuclei are in close relationship with the collectivity or deviations from a spherical shape. 
In this section, we discuss the $E2$ transition strengths of a few $|T_z|$ = 1/2 mirror pairs and $T$ = 1 triplets across the \textit{sd}-space. 
The calculated results corresponding to ISB \textit{sd}-shell interactions  DJ16A + CD + Av18 ($r_c$ = 0.55 fm) and USDCm are compared with the experimental data and isospin symmetric USDB and DJ16A results. The effective charges, $e_p$ = 1.36e and $e_n$ = 0.45e, are used to evaluate the $E2$ transition strengths and compared with the experimental data \cite{usd2}. {\color{black} Although these effective charges were considered for the isospin-symmetric case, we considered the same values for calculating $B(E2)$ within the ISB formalism. }
The calculated $E2$ transition strengths for $|T_z|$ = 1/2 nuclei from $A$ = 19-31 are shown in \autoref{tab:T_1_2}.
There are small changes seen in $E2$ transition strengths
for DJ16A and DJ16A(1, 2, 3, 4), so we keep the results of DJ16A and DJ16A(3) only in \autoref{tab:T_1_2}. Essentially for the DJ16A-based ISB interactions, the $B(E2)$ values for the DJ16A(4) or DJ16A + CD + Av18 ($r_c$ = 0.70 fm) interaction coincide with the DJ16A(3) or DJ16A + CD + Av18 ($r_c$ = 0.55 fm)  at 1$\%$ level. Hence, $E2$ transition strengths only for the DJ16A + CD + Av18 ($r_c$ = 0.55 fm) interaction are shown. {\color{black}The calculated $E2$ transition strengths for the  DJ16A + CD + Av18 ($r_c$ = 0.55 fm) interaction are consistent with the USDCm results except for the $^{25}$Al and $A$ = 29 pair. The $B(E2; 1/2^+_1 \to 5/2^+_1)$ of $^{25}$Al using USDCm shows slight overprediction compared to a underprediction for DJ16A + CD + Av18 ($r_c$ = 0.55 fm).} On ther other hand, the $B(E2; 3/2^+_1 \to 1/2^+_1)$ value of $^{29}$P using USDCm gives a significant overprediction compared to the DJ16A + CD + Av18 ($r_c$ = 0.55 fm) interaction. While $B(E2; 3/2^+_1 \to 1/2^+_1)$ of $^{29}$Si is underpredicted by the DJ16A + CD + Av18 ($r_c$ = 0.55 fm) interaction, the USDCm shows an overprediction for this transition. 
Similarly, in \autoref{tab:T_1}, the $E2$ transition strengths of $T$ = 1 triplets with $A$ = 18-30 are shown. The shell model results corresponding to the DJ16A + CD + Av18 ($r_c$ = 0.55 fm) interaction show that the $E2$ transition strengths are consistent with the USDCm results, except for $^{26}$Si and $A$ = 30 pair. 
The $\chi^2$ values for the interactions are denoted in \autoref{tab:T_1_2} and \autoref{tab:T_1}, where
\begin{equation}
\chi^2 =\sum_{i=1,N}(E_i -C_i)^2/Er_i^2
\end{equation}
with $E_i$ the experimental value, $Er_i$ the error in experiment, and $C_i$ the calculated value for nucleus $i$.
The $\chi^2$ values are smaller for the USDB and USDCm than for the DJ16A. One also notices that the CSB and CIB interactions reduce the value for the case of odd nuclei. 
{\color{black}However, these improvements are small, suggesting that the impact of ISB on $E2$ transition strength is minimal.}
Additionally, we performed another set of calculations for the DJ16A interaction using effective charges, $e_p$ = 1.202$e$ and $e_n$ = 0.196$e$ from Ref. \cite{sd_int3} that are obtained microscopically. The $\chi^2$ obtained for $|T_z|$ = 1/2 mirror pairs and $T$ = 1 triplets are 
 $\sqrt{\frac{1}{N}\chi^2}$} = 16.4 and 13.5, respectively, which are moderately worse than the earlier set of effective charges. 

\begin{table*}
\caption{The calculated $B(E2)$ values of $\left|T_z \right|=\frac{1}{2}$ $sd$-shell nuclei using different interactions, namely USDB, DJ16A, DJ16A + CD + Av18 ($r_c$ = 0.55 fm) and USDCm are compared to the experimental data. The effective charges are considered to be $e_p = 1.36e$ and $e_n = 0.45e$ in the calculations, and the $\chi^2$ value is denoted at the bottom of the table for each interaction.}
\label{tab:T_1_2}
\begin{tabular}{lcccccccc}
\hline
  & & & & &  \multicolumn{3}{c}{B(E2)$\downarrow$ [e$^2$fm$^4$]} & \\
 \hline \\[-7pt]
Isotope & $J^\pi_i$ & $J^\pi_f$ & Expt. & USDB & DJ16A & 
{\color{black} DJ16A+CD+Av18}&  USDCm  \\[+2pt]
 &  &  &  & &  & 
($r_c$ = 0.55 fm)&  \\[+2pt]

 \hline \\[-8pt]
$^{19}$Ne 	& $\frac{5}{2}^+_1$ 	& $\frac{1}{2}^+_1$ 	& 39.8 (15) \cite{ref:ENSDF}	& 36.9 & 39.0 & {\color{black}39.2} &  {\color{black} 37.2	}  	\\[+1pt]
$^{19}$F 		& $\frac{5}{2}^+_1$ 	& $\frac{1}{2}^+_1$ 	& 20.9 (2) \cite{ref:ENSDF}	 & 19.4 & 20.6 &  {\color{black}20.5}  	& {\color{black} 19.6}	  	\\[+1pt]
$^{21}$Na 	& $\frac{5}{2}^+_1$ 	& $\frac{3}{2}^+_1$ 	& 136.5 (92) \cite{ref:ENSDF} & 90.0 & 97.6& {\color{black}97.5} & {\color{black} 90.7}		\\[+1pt]
$^{21}$Ne		& $\frac{5}{2}^+_1$ 	& $\frac{3}{2}^+_1$ 	& 87.5 (58) \cite{ref:ENSDF} & 76.0 & 85.7&  {\color{black}85.9} &  {\color{black} 77.4	}	& \\[+1pt] 
$^{23}$Mg 	& $\frac{5}{2}^+_1$ 	& $\frac{3}{2}^+_1$ 	& 135 (15) \cite{ref:Henderson_21} &  117.2 & 124.4 &{\color{black}124.7} &  {\color{black} 117.8	}		\\[+1pt]
$^{23}$Na		& $\frac{5}{2}^+_1$ 	& $\frac{3}{2}^+_1$ 	& 106 (4) \cite{ref:Henderson_21} & 109.0 & 114.5 &  {\color{black}114.6} &  {\color{black} 109.4	} 	\\[+1pt] 
$^{25}$Al 		& $\frac{1}{2}^+_1$ 	& $\frac{5}{2}^+_1$ 	& 13.2 (3) \cite{ref:ENSDF} & 15.5 & 11.4 & {\color{black}11.1} &  {\color{black} 16.5	}  	\\[+1pt]
$^{25}$Mg	& $\frac{1}{2}^+_1$ 	& $\frac{5}{2}^+_1$ 	& 2.44 (4) \cite{ref:ENSDF} & 3.8 & 4.0 &  {\color{black}3.8} & {\color{black} 3.7} \\[+1pt] 
$^{27}$Si 		& $\frac{1}{2}^+_1$ 	& $\frac{5}{2}^+_1$ 	& 55.7 (64) \cite{ref:ENSDF} & 80.9 & 103.3 & {\color{black}103.4} &    {\color{black} 84.6} 	\\[+1pt]
$^{27}$Al		& $\frac{1}{2}^+_1$ 	& $\frac{5}{2}^+_1$ 	& 37.8 (11) \cite{ref:ENSDF} & 54.5 & 66.9&  {\color{black}66.9} &    {\color{black} 55.9} 	\\[+1pt] 
$^{29}$P 		& $\frac{3}{2}^+_1$ 	& $\frac{1}{2}^+_1$ 	& 14.3 (27) \cite{ref:ENSDF}& 45.7 & 28.7&  {\color{black}28.9} &    {\color{black} 44.8} 	\\[+1pt]
$^{29}$Si		& $\frac{3}{2}^+_1$ 	& $\frac{1}{2}^+_1$ 	& 21.7 (21) \cite{ref:ENSDF} & 31.0 & 15.8 &  {\color{black}15.4} &    {\color{black} 28.8}  	\\[+1pt] 
$^{31}$S 		& $\frac{3}{2}^+_1$ 	& $\frac{1}{2}^+_1$ 	& 40.5 (116) \cite{ref:ENSDF}& 39.6 & 48.1 &  {\color{black}48.7} &  {\color{black} 38.9} 	\\[+1pt]
$^{31}$P		& $\frac{3}{2}^+_1$ 	& $\frac{1}{2}^+_1$ 	& 24.3 (35) \cite{ref:ENSDF}	& 35.1 & 39.8 &  {\color{black} 39.9} &  {\color{black} 34.7}  	\\[+1pt]
\hline
& $\sqrt{\frac{1}{N}\chi^2}$ & &   & 11.1 & 13.1 & 12.1 & 10.8 \\[+1pt]
 \hline \\[-8pt]
\end{tabular}
\end{table*}

\begin{table*}
\centering
\begin{threeparttable}
\caption{The calculated $B(E2)$ values of $T$= 1 $sd$-shell nuclei using different interactions, namely USDB, DJ16A, DJ16A + CD + Av18 ($r_c$ = 0.55 fm) and USDCm are compared to the experimental data. The effective charges are considered to be $e_p = 1.36e$ and $e_n = 0.45e$ in the calculations, and all transitions are from 2$^+_1 (T = 1)$ to 0$^+_1 (T = 1)$ state. The $\chi^2$ value is denoted at the bottom of the table for each interaction where $^{18}$F result is excluded as it lacks experimental error information.}
\label{tab:T_1}
\begin{tabular}{lcccccc}
\hline
 & & & &\multicolumn{2}{c}{B(E2)$\downarrow$ [e$^2$fm$^4$]}\\
 \hline \\[-7pt]
Isotope 		& 	Expt. & USDB & DJ16A &
{\color{black} DJ16A+CD+Av18 } 
&   USDCm  		\\[+2pt]
		& 	 &  &  &
($r_c$ = 0.55 fm)
&    		\\[+2pt]
 \hline \\[-8pt]
 $^{18}$Ne 	&	49.6 (50) \cite{ref:ENSDF} & 29.8 & 29.9 &{\color{black}29.8} &  30.4 							\\[+1pt]
$^{18}$F & $>$ 22.4 \cite{ref:ENSDF} & 13.2 & 13.3 & 13.3 & 13.4
 \\[+1pt]
$^{18}$O & 	9.3 (3)	\cite{ref:ENSDF} &  3.3 & 3.3 &{\color{black}3.3} &  3.3  	 						\\[+1pt] 
$^{22}$Mg 	& 	76.2 (92) \cite{ref:Henderson_18} & 65.7 & 70.3 &  {\color{black}70.4} &   {\color{black} 66.3}  				\\[+1pt]
$^{22}$Na & 56 (19) \cite{ref:ENSDF}& 57.0 & 62.5 & 62.6 & 57.6
 \\[+1pt]
$^{22}$Ne		& 	46.9 (5) \cite{ref:Henderson_18, ref:ENSDF}$^\dagger$ & 48.9 & 55.2 &{\color{black}55.3} &  {\color{black} 49.9} 	\\[+1pt]
$^{26}$Si 	& 	70.0 (69) \cite{ref:ENSDF} &  47.0 & 69.4&{\color{black}69.4} & {\color{black} 48.3} 								  \\[+1pt]
$^{26}$Al & 58 (11) \cite{ref:ENSDF}& 57.4 & 74.7 & 75.9 & 58.7
 \\[+1pt]
$^{26}$Mg	& 	61.3 (26) \cite{ref:ENSDF} &  68.9 & 80.2 & {\color{black}79.8} &   {\color{black} 69.2}									\\[+1pt]
$^{30}$S 	& 	68.7 (40) \cite{ref:ENSDF} & 59.4 & 71.6 &{\color{black}70.4} &  {\color{black} 58.4}  	 								\\[+1pt]
$^{30}$P & 63.3 (39) \cite{ref:ENSDF} & 53.5 & 65.2 & 58.2 & 54.7
 \\[+1pt]
$^{30}$Si	& 	49.9 (65) \cite{ref:ENSDF} & 47.9 & 59.0& {\color{black}58.4} &  {\color{black} 48.4} 									\\[+1pt]
\hline
& $\sqrt{\frac{1}{N}\chi^2}$  & 6.5  & 8.3 & 8.3 & 6.6\\[+1pt]
\hline
 
\end{tabular}
\begin{tablenotes}
\item $\dagger$ - Weighted average of values in  Ref.~\cite{ref:ENSDF} and Ref.~\cite{ref:Henderson_18}
\end{tablenotes}
\vspace{-10pt}
\end{threeparttable}
\end{table*}

We further investigated the isoscalar ($M_0$) and isovector ($M_1$) matrix elements for the $E2$ transition strengths reported in \autoref{tab:T_1_2} and \autoref{tab:T_1} for the 
 effective charges, $e_p = 1.36e$ and $e_n =045e$, corresponding to the ISB interactions (DJ16A + CD + Av18 ($r_c$ = 0.55 fm) and USDCm), and compared them with the experimental data following the formalism of Ref. \cite{brown_iso}. The $M_0$ and $M_1$ matrix elements are defined as
\begin{equation}
    M_0 = \frac{\sqrt{B(E2; T_z < 0)} + \sqrt{B(E2; T_z > 0)} }{2}
\end{equation}

and 
\begin{equation}
    M_1 = \frac{\sqrt{B(E2; T_z < 0)} - \sqrt{B(E2; T_z > 0)} }{\Delta T_z}.
\end{equation}

The ratios of experimental to calculated isoscalar ($M_0$) and isovector ($M_1$) matrix elements are shown in \autoref{BE2_M0}. 
The figures on the left show the ratios of $M_0$ for DJ16A + CD + Av18 ($r_c$ = 0.55 fm) (top-left panel) and USDCm (bottom-left panel). On the other hand, the figures on the right show the same for $M_1$ using the same sets of interactions. The figure shows that the calculated isoscalar matrix elements exhibit small deviations from the experimental $M_0$ matrix elements compared to the isovector matrix elements. 
Also, among the $M_0$ matrix elements, the USDCm results show slightly more deviation from the experimental data compared to the DJ16A + CD + Av18 ($r_c$ = 0.55 fm) results, particularly for the $A$ = 29 pair.} On the other hand, the calculated $M_1$ matrix elements for the $A$ = 18, and 19 pairs were close to the experimental matrix elements; however, other pairs showed significant deviations. 
In fact, the $M_1$ matrix element obtained for the USDCm interaction corresponding to $B(E2; 2^+_1 \to 0^+_1)$ transitions of {\color{black}the} $A$ = 26 pair reported in \autoref{tab:T_1} is almost {\color{black}2.5} times larger than the experimental $M_1$ matrix element with a large error bar. 

\begin{figure*}
    \includegraphics[scale = 0.438]{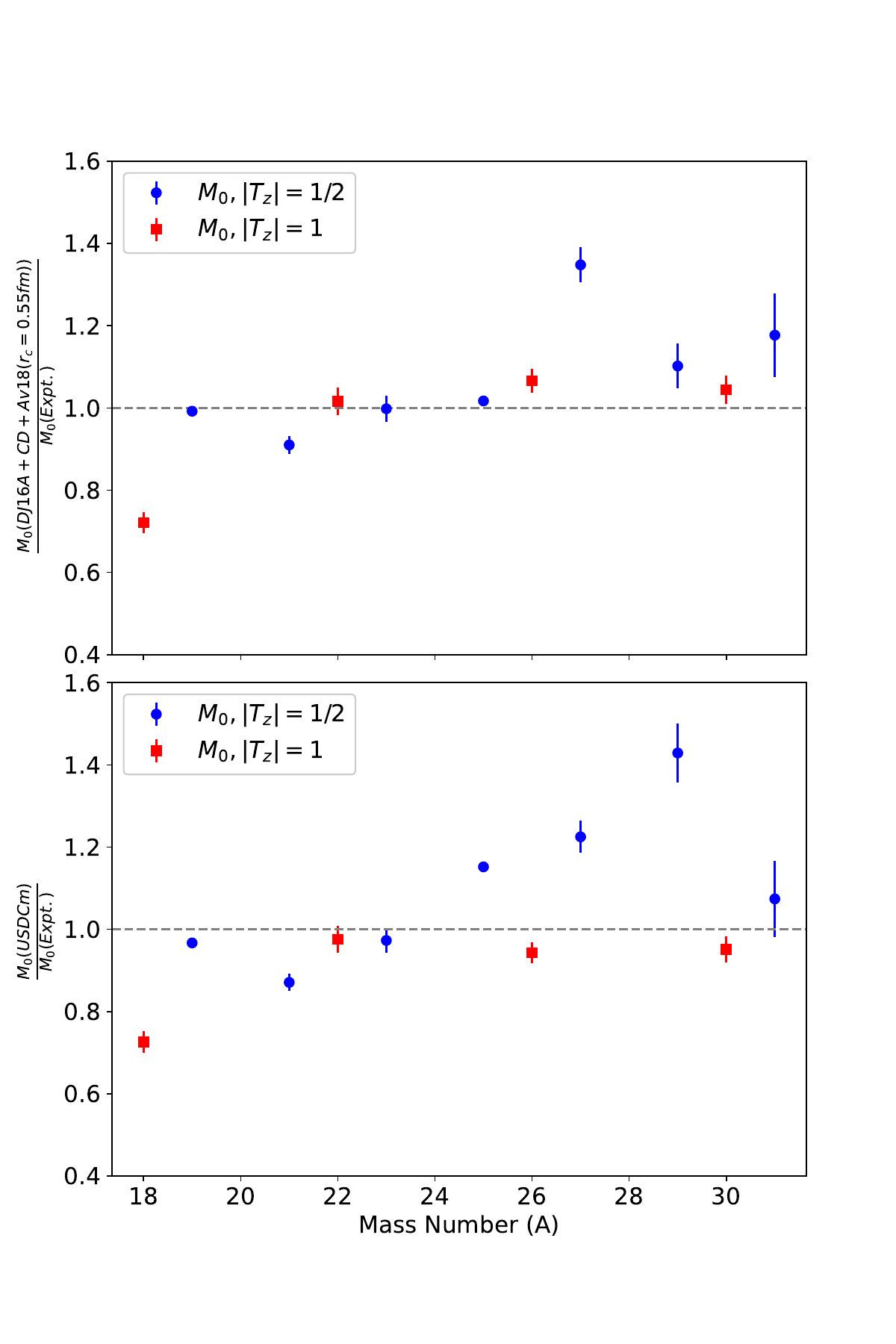}
    \includegraphics[scale = 0.438]{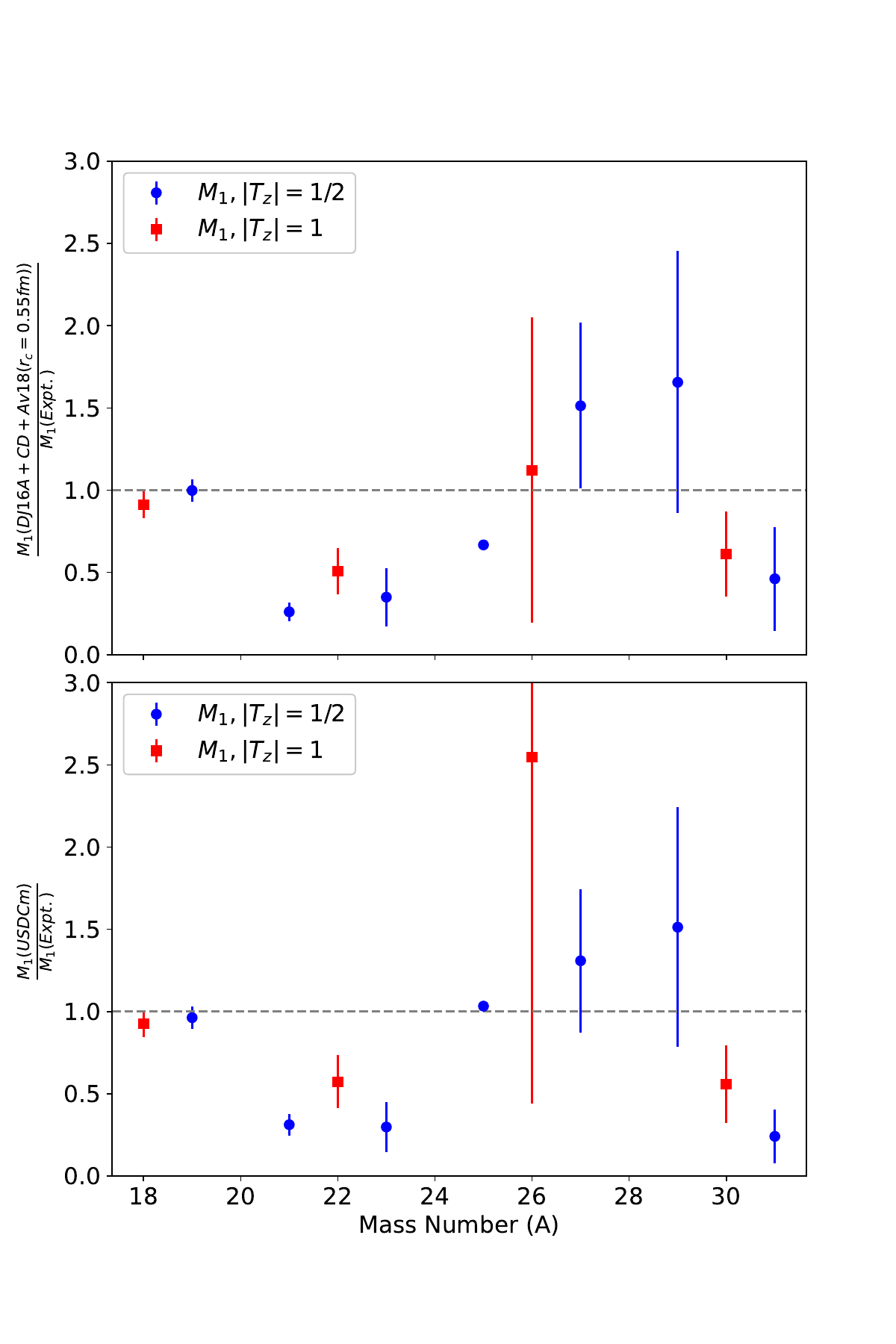}
    \caption{ (Left  panels) The ratios of experimental to calculated $M_0$ are shown for DJ16A + CD + Av18 ($r_c$ = 0.55 fm) (top-left panel) and USDCm (bottom-left panel); (right panels) ratios of experimental to calculated $M_1$ are shown for DJ16A + CD + Av18 ($r_c$ = 0.55 fm) (top-right panel) and USDCm (bottom-right panel)}
    \label{BE2_M0}
\end{figure*}

Now, we focus on the ratios of $M_0$ and $M_1$ matrix elements corresponding to $|T_z|$ = 1/2 pair: $A$ = 29 and $|T_z|$ = 1 pair $A$ = 26, for which at least one of the $E2$ transition strengths for DJ16A + CD + Av18 ($r_c$ = 0.55 fm) interaction is far away from the USDCm results. 
From \autoref{BE2_M0}, it can be seen that the $M_0$ matrix elements of the $A$ = 26 pair corresponding to 
{\color{black} DJ16A+CD+Av18 ($r_c$ = 0.55 fm)}
are slightly higher than the USDCm matrix element. On the other hand, the $M_1$ matrix element for USDCm is much higher than the 
{\color{black} DJ16A+CD+Av18 ($r_c$ = 0.55 fm)} interaction. Finally, the calculated $M_1$ matrix elements of {\color{black}the} $A$ = 29 pair are almost the same for both ISB interactions, and the difference in $E2$ transition strength between USDCm and DJ16A-based interaction {\color{black}is} isoscalar in nature.

\section{Summary and Conclusions} 
\label{sect 4}
In this work, we have developed and tested the ISB versions of the DJ16A interaction. {\color{black}Starting from the isospin-symmetric version of this interaction from Ref. \cite{dj16_6A}, we added Coulomb-CD and also the  CSB and CIB effects from realistic Av18 interaction at two different radial cut-offs are added to it. Finally, we considered two sets of interactions : (i) DJ16A + CD + Av18 ($r_c$ = 0.55 fm), and (ii) DJ16A + CD + Av18 ($r_c$ = 0.70 fm), and their proton SPEs are modified to reproduce experimental $b~(A = 17)$. We then employed both of these DJ16A-based interactions along with the USDCm to calculate the $b$-parameters of $|T_z|$ = 1/2 and $|T_z|$ = 1 nuclei across the $sd$-shell and compared them with the experimental data. 
Based on the $b_{rms}$ values of the g.s. of $|T_z|$ = 1/2 and $|T_z|$ = 1 nuclei across $sd$-shell, the performance of both DJ16A-based interactions are comparable to the USDCm results for nuclei with $A<$30. Additionally, we calculated the $c$-parameters for the $T$ = 1 nuclei across the $sd$-shell, and the calculated $c$-parameters using both DJ16A-based interactions along with the USDCm interaction can reproduce the staggering pattern of the experimental data correctly. The $c_{rms}$ suggests that DJ16A + CD + Av18 ($r_c$ = 0.55 fm) results show slightly more deviations than the other two ISB interactions.

We then used these interactions to calculate MEDs in the low-energy spectra of a few nuclear mirror pairs around $A$ = 20 and compared them to the recently developed USDCm results and experimental data. For this work, $|T_z|$ = 1/2 pairs: ($^{19}$Ne - $^{19}$F) and ($^{21}$Na - $^{21}$Ne) and $|T_z|$ = 1 pairs: ($^{18}$Ne - $^{18}$O) and ($^{22}$Mg - $^{22}$Ne) were considered. The calculated MEDs are found to be consistent with the experimental data within certain error ranges for the lowest two or three excited states, {\color{black} specifically for the $|T_z|$ = 1 pairs}. For these states, the order of energy levels agrees with that of the experimental data. The TEDs for low-lying states of $T$ = 1 triplets in the $A$ = 18 and 22 systems were also studied, and the calculated results of the DJ16A + CD + Av18 ($r_c$ = 0.70 fm) interaction are found to be consistent with the experimental data.

Finally, we calculated the $E2$ transition strengths of some $|T_z|$ = 1/2 mirror pairs and $T = 1$ triplets
using the DJ16A + CD + Av18 ($r_c$ = 0.55 fm) interaction along with the USDCm interaction, and compared them with the experimental data. The shell model results are, on average, in good agreement with the experimental data. Almost the same $B(E2)$ values for both DJ16A + CD + Av18 ($r_c$ = 0.55 fm) and DJ16A + CD + Av18 ($r_c$ = 0.70 fm) suggest that the effects of the difference in CSB and CIB parts on $E2$ transition strengths are quite small. 
Because of this, only DJ16A + CD + Av18 ($r_c$ = 0.55 fm) results are reported in this work. Further, we calculated the ratios of theoretical to experimental isoscalar ($M_0$) and isovector ($M_1$) components for those transitions corresponding to DJ16A + CD + Av18 ($r_c$ = 0.55 fm) and USDCm interactions. While the $M_0$ ratios are close to the unity line for both interactions, the $M_1$ ratios show significant deviations in some cases. 

From the above discussion, it can be concluded that both DJ16A-based ISB interactions yield comparable results for the $b$-parameters, MEDs, and $E2$ transition strengths. A slight difference is observed in the $c$-parameters and TEDs, where the DJ16A + CD + Av18 interaction with $r_c$ = 0.70 fm shows improved performance. Although some deviations from experimental data persist, the overall description of $b$- and $c$-parameters, MEDs of mirror pairs, and TEDs of $T$ = 1 triplets is satisfactory within the present framework. This agreement is achieved by incorporating short-range correlations in the CSB and CIB components through a radial cutoff in the physically reasonable range $r_c$ = 0.5–0.7fm \cite{SIA1974, ACS1983, BBW1977}. 
{\color{black} Overall, both DJ16A-based ISB interactions provide an improved description of several observables associated with ISB effects, although their performance varies across the different quantities considered.
}
In contrast, the USDCm interaction, being phenomenologically fitted with a larger number of adjustable parameters, achieves closer agreement with experimental data for most observables.

\section*{ACKNOWLEDGMENTS}
 We acknowledge financial support from research Grant No. ANRF/ARGM/2025/001130/TS from Anusandhan National Research Foundation [ANRF], India. CS acknowledges UK STFC grant ST/Y000358/1. 
 N.S. acknowledges the KAKENHI budgets 25K00995, 25H01268, 25K07294,
25K07330, and 24H00239. The shell model calculations were performed using the KSHELL code \cite{KShell}.



\begin{thebibliography}{44}
	\expandafter\ifx\csname natexlab\endcsname\relax\def\natexlab#1{#1}\fi
	\expandafter\ifx\csname bibnamefont\endcsname\relax
	\def\bibnamefont#1{#1}\fi
	\expandafter\ifx\csname bibfnamefont\endcsname\relax
	\def\bibfnamefont#1{#1}\fi
	\expandafter\ifx\csname citenamefont\endcsname\relax
	\def\citenamefont#1{#1}\fi
	\expandafter\ifx\csname url\endcsname\relax
	\def\url#1{\texttt{#1}}\fi
	\expandafter\ifx\csname urlprefix\endcsname\relax\def\urlprefix{URL }\fi
	\providecommand{\bibinfo}[2]{#2}
	\providecommand{\eprint}[2][]{\url{#2}}
	
	
	
	
	
	
	
\bibitem{Heisenberg}W. Heisenberg, \href{https://doi.org/10.1007/BF01342433}{Zeitschrift für Physik {\bf77}, 1-11 (1932)}.

\bibitem{Wigner} E. Wigner, ``On the consequences of the symmetry of the nuclear hamiltonian on the spectroscopy of nuclei'', \href{https://doi.org/10.1103/PhysRev.51.106}{Phys. Rev. \textbf{51}, 106 (1937).}

\bibitem{brown_89} W. E. Ormand and B. A. Brown, ``Empirical isospin-nonconserving hamiltonians for shell-model calculations'', \href{https://doi.org/10.1016/0375-9474(89)90203-0}{Nucl. Phys. A \textbf{491}, 1 (1989).}

\bibitem{brown_85} W. E. Ormand and B. A. Brown, ``Calculated isospin-mixing corrections to Fermi $\beta$-decays in 1s0d-shell nuclei with emphasis on A = 34'', \href{https://doi.org/10.1016/0375-9474(85)90341-0}{Nucl. Phys. A \textbf{440}, 274 (1985).}


\bibitem{ISB_rev} N. A. Smirnova, Isospin-symmetry breaking within the nuclear shell model: present status and developments, 
        \href{https://doi.org/10.3390/physics5020026}
	{Physics {\bf 5}, 352–380 (2023).}

\bibitem{ISB_sakshi} S. Shukla, P.C. Srivastava and K. Kaneko, Shell model study of isobaric analog states for $T_z$ = $\pm$ 2 nuclei using isospin non-conserving interactions, 
\href{https://doi.org/10.1088/1361-6471/ae2eec}
{J. Phys. G: Nucl. Part. Phys.  \textbf{53}, 015103 (2026).}
    

\bibitem{DFT} P. Baczyk, J. Dobaczewski, M. Konieczka, W. Satu\l{}a, T. Nakatsukasa, and K. Sato, ``Isospin-symmetry breaking in masses of N = Z nuclei'', \href{https://doi.org/10.1016/j.physletb.2017.12.068}{Phys. Lett. B \textbf{778}, 178 (2018).}

\bibitem{ISB_rev2} J. A. Sheikh, S. P. Rouoof, R. N. Ali, N. Rather, C. Sarma, and P. C. Srivastava, ''Isospin symmetry breaking in atomic nuclei'', \href{https://doi.org/10.3390/sym16060745}{Symmetry {\bf} 16, 745 (2024)}.

\bibitem{ISB_beta1}I.S.Towner, ``Mirror asymmetry in allowed gamow-teller $\beta$-decay'', \href{https://doi.org/10.1016/0375-9474(73)90172-3}{Nucl. Phys. A \textbf{216}, 589 (1973).}

\bibitem{ISB_beta2} N. A. Smirnova and C. Volpe, ``On the asymmetry of Gamow–Teller $\beta$-decay rates in mirror nuclei in relation with second-class currents'', \href{https://doi.org/10.1016/S0375-9474(02)01392-1}{Nucl. Phys. A \textbf{714}, 441 (2003).}

\bibitem{ISB_effect1} B. Blank and M. J. G. Borge, ``Nuclear structure at the proton drip line: Advances with nuclear decay studies'', \href{https://doi.org/10.1016/j.ppnp.2007.12.001}{Prog. Part. Nucl. Phys. \textbf{60}, 403 (2008).}

\bibitem{ISB_effect2} E. Hagberg, V. T. Koslowsky, J. C. Hardy, I. S. Towner, J. G. Hykawy, G. Savard, and T. Shinozuka, ``Tests of Isospin Mixing Corrections in Superallowed 0$^+$ $\to$ 0$^+$ $\beta$ Decays'', \href{https://doi.org/10.1103/PhysRevLett.73.396}{Phys. Rev. Lett. \textbf{73}, 396 (1994).}

\bibitem{ISB_effect3} E. Farnea et al., ``Isospin mixing in the N=Z nucleus $^{64}$Ge'', \href{https://doi.org/10.1016/S0370-2693(02)03022-8}{Phys. Lett. B \textbf{551}, 56 (2003).}

\bibitem{ISB_pf} A. P. Zuker, S. M. Lenzi, G. Martínez-Pinedo, and A. Poves, ``Isobaric Multiplet Yrast Energies and Isospin Nonconserving Forces'', \href{https://doi.org/10.1103/PhysRevLett.89.142502}{Phys. Rev. Lett. \textbf{89}, 142502 (2002).}

\bibitem{ISB_18Ne} D.W.Bardayan \textit{et al.}, ``Observation of the Astrophysically Important 3$^+$ State in $^{18}$Ne via Elastic Scattering of a Radioactive $^{17}$F Beam from $^1$H'', \href{https://doi.org/10.1103/PhysRevLett.83.45}{Phys. Rev. Lett. \textbf{83}, 45 (1999).}

\bibitem{TE_1}R. G. Thomas, An Analysis of the Energy Levels of the Mirror Nuclei, C$^{13}$ and N$^{13}$, \href{https://doi.org/10.1103/PhysRev.88.1109}{Phys. Rev. {\bf88}, 1109 (1952)}.

\bibitem{TE_2}J. B. Ehrman, On the Displacement of Corresponding Energy Levels of C$^{13}$ and N$^{13}$ \href{https://doi.org/10.1103/PhysRev.81.412}{Phys. Rev. {\bf 81}, 412 (1951)}.

\bibitem{ISB_rev3} M.A. Bentley, S.M. Lenzi, ``Coulomb energy differences between high-spin states in isobaric multiplets'', \href{https://doi.org/10.1016/j.ppnp.2006.10.001}{Prog. Part. Nucl. Phys. \textbf{59}, 497 (2007).}

\bibitem{MED_A20} C. Yuan, C. Qi, F. Xu, T. Suzuki, and T. Otsuka, Mirror energy difference and the structure of loosely bound proton-rich nuclei around $A$ = 20,
\href{https://doi.org/10.1103/PhysRevC.89.044327}{Phys. Rev. C. {\bf 89}, 044327 (2014).}

\bibitem{isb_sd_2013} Y. H. Lam, N. A. Smirnova, and E. Caurier, ''Isospin nonconservation in \textit{sd}-shell nuclei'', \href{http://dx.doi.org/10.1103/PhysRevC.87.054304}{Phys. Rev. C \textbf{87}, 054304 (2013).}

\bibitem{usd1} B. H. Wildenthal, ``Empirical strengths of spin operators in nuclei'', \href{https://doi.org/10.1016/0146-6410(84)90011-5}{Prog. Part. Nucl. Phys. \textbf{11}, 5 (1984).}

\bibitem{usd2} B. A. Brown and B. H. Wildenthal, ``Status of the Nuclear Shell Model'', \href{https://doi.org/10.1146/annurev.ns.38.120188.000333}{Annu. Rev. Nucl. Part. Sci. \textbf{38}, 29 (1988).}
 
\bibitem{usdb} B. A. Brown and W. A. Richter, ``New ``USD'' Hamiltonians for the $sd$ shell'', \href{http://dx.doi.org/10.1103/PhysRevC.74.034315}{Phys. Rev. C {\bf 74}, 034315 (2006).}

\bibitem{usdc} A. Magilligan and B. A. Brown, ``New isospin-breaking “USD” Hamiltonians for the \textit{sd} shell'', \href{https://doi.org/10.1103/PhysRevC.101.064312}{Phys. Rev. C \textbf{101}. 064312 (2020).}


\bibitem{brown_correction} There is a misprint in Table II of this reference. The matrix element labeled v(5 3 3 3 : 1 0) should be labeled v(3 5 3 3 : 1 0) (B.A.Brown, private communication).


\bibitem{subrajit_M1} S. Sahoo and P. C. Srivastava, ``Isoscalar, isovector and orbital contributions in \textit{M1} transitions from analogous \textit{ M1} and
Gamow-Teller transitions in T = 1/2 mirror nuclei'', \href{https://doi.org/10.1103/PhysRevC.110.054305}{Phys. Rev. C \textbf{110}, 054305 (2024)}. 

\bibitem{ISB_abinitio} M. S. Martin , S. R. Stroberg, J. D. Holt, and K. G. Leach, ``Testing isospin symmetry breaking in \textit{ab initio} nuclear theory'', \href{https://doi.org/10.1103/PhysRevC.104.014324}{Phys. Rev. C. \textbf{104}, 014324 (2022).}

\bibitem{MED_sd1} H. H. Li, J. G. Li, M. R. Xie, and W. Zuo, \textit{Ab initio} calculations of mirror energy difference in $sd$-shell nuclei, \href{https://doi.org/10.1088/1674-1137/acf035}{Chinese Phys. C {\bf 47}, 124101 (2023)}.

\bibitem{MED_sd2} H. H. Li, Q. Yuan, J. G. Li, M. R. Xie, S. Zhang, Y. H. Zhang, X. X. Xu, N. Michel, F. R. Xu, and W. Zuo, Investigation of isospin-symmetry breaking in mirror energy difference and nuclear mass with \textit{ab initio} calculations, \href{https://doi.org/10.1103/PhysRevC.107.014302}{Phys. Rev. C {\bf 107}, 014302 (2023).}

\bibitem{ref:Henderson_21} J. Henderson, G. Hackman,\textit{ et. al}., ``Coulomb excitation of the $|T_z|$ = 1/2 , \textit{A} = 23 mirror pair'', \href{https://doi.org/10.1103/PhysRevC.105.034332}{Phys. Rev. C \textbf{105}, 034332 (2022).}

\bibitem{stroberg_2022} S. R. Stroberg, J. Henderson, G. Hackman, P. Ruotsalainen, G. Hagen, and J. D. Holt, ``Systematics of \textit{E2} strength in the \textit{sd }shell with the valence-space in-medium
similarity renormalization group", \href{https://doi.org/10.1103/PhysRevC.105.034333}{Phys. Rev. C \textbf{105}, 034333 (2022)}

\bibitem{38Ca_2023}  T. Beck, A. Gade, B. A. Brown, \textit{et al.}, ``Probing proton cross-shell excitations through the two-neutron removal from $^{38}$Ca'', \href{https://doi.org/10.1103/PhysRevC.108.L061301}{Phys. Rev. C. \textbf{108}, L061301 (2023).}

\bibitem{28S_2022} S. A. Gillespie, K. W. Brown, R. J. Charity, \textit{et al.} , ``Proton decay spectroscopy of $^{28}$S and $^{30}$Cl'', \href{https://doi.org/10.1103/PhysRevC.105.044321}{Phys. Rev. C. \textbf{105}, 044321 (2022).}

\bibitem{22Al_2024} S. E. Campbell, G. Bollen, B. A. Brown, \textit{et al.}, ``Precision Mass Measurement of the Proton Dripline Halo Candidate $^{22}$Al'', \href{https://doi.org/10.1103/PhysRevLett.132.152501}{Phys. Rev. Lett. \textbf{132}, 152501 (2024).}

\bibitem{22Si_2023} J. G. Li, H. H. Li, S. Zhang, Y. M. Xing, and W. Zuo. ``Double-magicity of proton drip-line nucleus $^{22}$Si with \textit{ab initio} calculation'', \href{https://doi.org/10.1016/j.physletb.2023.138197}{Phys. Lett. B \textbf{846}, 138197 (2023).}







\bibitem{Na_NCSM} C. Sarma, and P. C. Srivastava, ``Nuclear structure study of $^{20-23}$Na isotopes with \textit{ab initio} no-core shell-model'', \href{https://doi.org/10.1007/s00601-025-01995-9
}{Few-Body Syst \textbf{66}, 23 (2025)}



\bibitem{double_ols}  A.F. Lisetskiy, B.R. Barrett, M.K.G. Kruse, P. Navrátil, I. Stetcu,
J.P. Vary, ``\textit{Ab-initio} shell model with a core, \href{https://doi.org/10.1103/PhysRevC.78.044302}{Phys. Rev. C \textbf{78}, 044302
(2008)}. 
 
\bibitem{sd_int1} E. Dikmen, A. F. Lisetskiy, B. R. Barrett, P. Maris, A.M. Shirokov, J.P. Vary, ``$Ab$ $initio$ effective interactions for sd-shell valence nucleons'', \href{https://doi.org/10.1103/PhysRevC.91.064301}{Phys. Rev. C {\bf91}, 064301 (2015).} 

\bibitem{dj16_6A} I. J. Shin , N. A. Smirnova, A M. Shirokov , Z. Yang, B. R. Barrett, Z. Li ,
Y. Kim, P. Maris, and J. P. Vary, ``Improving the \textit{sd}-shell effective interaction obtained from
the Daejeon16 nucleon-nucleon interaction'', \href{https://doi.org/10.1103/PhysRevC.110.034306}{Phys. Rev. C \textbf{110}, 034306 (2024).} 



\bibitem{Kuo65} T. T. S. Kuo and G. E. Brown, ``Structure of finite nuclei and the free nucleon-nucleon interaction: An application to $^{18}$O and $^{18}$F'', \href{https://doi.org/10.1016/0029-5582(66)90131-3}{Nucl. Phys. \textbf{85}, 40 (1966).}

\bibitem{Kuo68} T. T. S. Kuo and G. E. Brown, ``Reaction matrix elements for the 0f-1p shell nuclei'', \href{https://doi.org/10.1016/0375-9474(68)90353-9}{Nucl. Phys. \textbf{114}, 241 (1968).}

\bibitem{ckpot} S. Cohen and D. Kurath, Effective interactions for the 1p shell, \href{https://doi.org/10.1016/0029-5582(65)90148-3}{Nucl. Phys. {\bf 73}, 1 (1965).}

\bibitem{kb3g} A. Poves, J. Sanchez-Solano, E. Caurier, and F. Nowacki, ``Shell model study of the isobaric chains \textit{A} = 50,
\textit{A} = 51 and \textit{A }= 52 '', \href{https://doi.org/10.1016/S0375-9474(01)00967-8}{Nucl. Phys. A \textbf{694}, 157 (2001).}

\bibitem{gxpf1a} M. Honma, T. Otsuka, B. A. Brown, and T. Mizusaki, ``New effective interaction for \textit{pf}-shell nuclei and its implications for the stability of the \textit{N} = \textit{Z} = 28 closed core'', \href{https://doi.org/10.1103/PhysRevC.69.034335}{Phys. Rev. C. \textbf{69}, 034335 (2004).}


    \bibitem{Otsuka1} T. Otsuka, A. Gade, O. Sorlin, T. Suzuki, and Y. Utsuno, ``Evolution of shell structure in exotic nuclei",
            \href{https://doi.org/10.1103/RevModPhys.92.015002}
            {Rev. Mod. Phys. {\bf 92}, 015002 (2020).}

     \bibitem{Caurier} E. Caurier, G. M-Pinedo, F. Nowacki, A. Poves, and A. P. Zuker, ``The shell model as a unified view of nuclear structure",
            \href{https://doi.org/10.1103/RevModPhys.77.427}
            {Rev. Mod. Phys. {\bf 77}, 427 (2005).}


\bibitem{ncsm} B. R. Barrett, P. Navrátil, and J. P. Vary, ``\textit{Ab initio} no core shell model'', \href{https://doi.org/10.1016/j.ppnp.2012.10.003}{Prog. Part. Nucl. Phys. \textbf{69}, 131 (2013).}


\bibitem{vsimsrg1} H. Hergert, S. Binder, A. Calci, J. Langhammer, and R. Roth, ``\textit{Ab Initio} Calculations of Even Oxygen Isotopes with Chiral Two-Plus-Three-Nucleon Interactions'', \href{https://doi.org/10.1103/PhysRevLett.110.242501}{Phys.Rev.Lett. \textbf{110}, 242501 (2013).}

\bibitem{vsimsrg2} S. K. Bogner, H. Hergert, J. D. Holt, A. Schwenk, S. Binder, A. Calci, J. Langhammer, and R. Roth, ``Nonperturbative Shell-Model Interactions from the In-Medium Similarity Renormalization Group'', \href{https://doi.org/10.1103/PhysRevLett.113.142501}{Phys. Rev. Lett. \textbf{113}, 142501 (2014).}

\bibitem{cc} G. R. Jansen, J. Engel, G. Hagen, P. Navratil, and A. Signoracci, ``Ab Initio Coupled-Cluster Effective Interactions for the Shell Model: Application to Neutron-Rich Oxygen and Carbon Isotopes'', \href{https://doi.org/10.1103/PhysRevLett.113.142502}{Phys. Rev. Lett. \textbf{113}, 142502 (2014).}




\bibitem{priyanka1} P. Choudhary, P. C. Srivastava, and P. Navr\'atil, ``$Ab$ $initio$ no-core shell model study of $^{10-14}$B
isotopes with realistic $NN$ interactions", 
\href{https://doi.org/10.1103/PhysRevC.102.044309}
{Phys. Rev. C {\bf 102}, 044309 (2020).}

\bibitem{priyanka2} P. Choudhary, P. C. Srivastava, M. Gennari, and P. Navr\'atil, \textit{Ab initio} no-core shell-model description of $^{10-14}$C isotopes, 
\href{https://doi.org/10.1103/PhysRevC.107.014309}
{Phys. Rev. C  {\bf 117}, 014309 (2023).}


\bibitem{arch2} A. Saxena and P. C. Srivastava, \textit{Ab initio} no-core shell model study of $^{18-23}$O and $^{18-24}$F isotopes, \href{https://doi.org/10.1088/1361-6471/ab6f1d}{J. Phys. G: Nucl. Part. Phys. {\bf 47}, 055113 (2020).}

\bibitem{chandan} C. Sarma and P. C. Srivastava, \textit{Ab-initio} no-core shell model study of $^{18-24}$Ne isotopes, 
\href{https://doi.org/10.1088/1361-6471/acb962}
{J. Phys. G: Nucl. Part. Phys. {\bf 50}, 045105 (2023).}

\bibitem{dj16_4A} 
N. A. Smirnova, B. R. Barrett, Y. Kim, I. J. Shin, A. M. Shirokov, E. Dikmen, P. Maris, and J. P. Vary, ``Effective interactions in the $sd$ shell'', \href{https://doi.org/10.1103/PhysRevC.100.054329}{Phys. Rev. C {\bf 100}, 054329 (2019).}


\bibitem{jisp16} A. Shirokov, A. Mazur, J. Vary, and T. Weber, ``Realistic nuclear Hamiltonian: $Ab$ $exitu$ approach'', 
\href{https://doi.org/10.1016/j.physletb.2006.10.066}{Phys. Lett. B 644, 33 (2007).}

\bibitem{n3lo} D. R. Entem and R. Machleidt, ``Accurate charge-dependent nucleon-nucleon potential at fourth order of chiral perturbation theory'',
\href{https://doi.org/10.1103/PhysRevC.68.041001}{Phys. Rev. C 68, 041001(R) (2003).}

\bibitem{daejeon16} A. M. Shirokov, I. J. Shin, Y. Kim, M. Sosonkina, P. Maris, and J. P. Vary, ``N3LO NN interaction adjusted to light nuclei in $ab$ $exitu$ approach''
\href{https://doi.org/10.1016/j.physletb.2016.08.006}{Phys. Lett. B 761, 87 (2016).}

\bibitem{ols_1994} K. Suzuki and R. Okamoto, ``Effective interaction theory
and unitary-model-operator approach to nuclear saturation problem'', \href{https://doi.org/10.1143/ptp/92.6.1045}{Prog. Theor. Phys. {\bf 92}, 1045 (1994).}



\bibitem{priyanka_EPJA}
P. Choudhary and P.C. Srivastava,
``Study of S, Cl and Ar isotopes with $N \geq Z$  using microscopic effective sd-shell interactions'', \href{https://doi.org/10.1140/epja/s10050-023-01013-8}{Eur. Phys. J. A {\bf 59}, 97 (2023).}



\bibitem{priyanka_NPA} P. Choudhary and  P. C. Srivastava, 
``Nuclear structure properties of Si and P isotopes with the microscopic effective interactions'', \href{ https://doi.org/10.1016/j.nuclphysa.2023.122629}{Nucl Phys. A {\bf 1033}, 122629  (2023).}



\bibitem{subhrajit_NPA} S. Sahoo, P. C. Srivastava, and T. Suzuki, ''Study of structure and radii for $^{20-31}$Na isotopes using microscopic interactions'', \href{https://doi.org/10.1016/j.nuclphysa.2023.122618}{Nucl. Phys. A {\bf 1032}, 122618 (2023).}

\bibitem{chandan_PRC} C. Sarma, O. D. Matteo, A. Abhishek, and P. C. Srivastava, ``Prediction of the neutron drip line in oxygen isotopes using quantum computation'', \href{https://doi.org/10.1103/PhysRevC.108.064305}{Phys. Rev. C \textbf{108}, 064305 (2023).}


\bibitem{sd_int3} Z. Li, N. A. Smirnova, A. M. Shirokov, I. J. Shin, B. R. Barrett, P. Maris, J. P. Vary, ``Effective operators for valence space calculations from the {\itshape ab initio} No-Core Shell Model'', \href{https://doi.org/10.1142/9789819813353_0011}{Akito Arima Memorial Volume, World Scientific Publishing Co. Pte. Ltd. }



\bibitem{NS} J. A. Nolen, J.P. Schiffer,  ``Coulomb Energies'', \href{https://doi.org/10.1146/annurev.ns.19.120169.002351}{Ann. Rev. Nucl. Sci. \textbf{1969}, 19, 471–526.}


\bibitem{Suzuki92} T. Suzuki, H. Sagawa, and A. Arima, ``Effects of valence nucleon orbits and charge symmetry breaking interaction on the Nolen-Schiffer anomaly of mirror nuclei'', \href{https://doi.org/10.1016/0375-9474(92)90250-N}{Nucl. Phys. A \textbf{536}, 141 (1992)}.

\bibitem{AV18_int} R. B. Wiringa, V. G. J. Stoks, and R. Schiavilla, ``Accurate nucleon-nucleon potential with charge-independence breaking'', \href{https://doi.org/10.1103/PhysRevC.51.38}{Phys. Rev. C. \textbf{51}, 38 (1995).}

\bibitem{AB1977}
M. R. Anastasio and G. E. Brown,
``Meson exchange and the theory of finite Fermi systems'',
\href{https://doi.org/10.1016/0375-9474(77)90650-9}{Nucl. Phys. \textbf{A285}, 516 (1977)}

\bibitem{BBW1977}
G. E. Brown, S.-O. Backman, E. Oset, and W. Weise,
``Calculation of spin-dependent parameters in the Landau-Migdal theory of nuclei'',
\href{https://doi.org/10.1016/0375-9474(77)90017-3}{Nucl. Phys. \textbf{A286}, 191 (1977)}

\bibitem{Atti2015}
Claudio Ciofi degli Atti,
``In-medium short-range dynamics of nucleons:
Recent theoretical and experimental advances'',
\href{http://dx.doi.org/10.1016/j.physrep.2015.06.002}
{Physics Reports \textbf{590}, 1 (2015).}

\bibitem{Kaneko2017} K. Kaneko, Y. Sun, T. Mizusaki, S. Tazaki, and S. K. Ghorui, 
``Isospin-symmetry breaking in superallowed Fermi $\beta$-decay due to isospin-nonconserving forces'',
\href{http://dx.doi.org/10.1016/j.physletb.2017.08.056} {Phys. Lett. B \textbf{773}, 521 (2017).}

\bibitem{bohr} A. Bohr and B. R. Mottelson, \textit{Nuclear Structure}, Vol. I (World Scientific,  1997).

\bibitem{TES_A16_PLB} K. Ogawa, H. Nakada, S. Hino, and R. Motegi, ``Thomas-Ehrman shifts in nuclei around $^{16}$O and role of residual nuclear interaction'', \href{https://doi.org/10.1016/S0370-2693(99)00993-4}{Phys. Lett. B \textbf{464}, 157 (1999).}

\bibitem{git_repo} GitHub repository, \href{https://github.com/chandan-sarma/ISB_interaction}{https://github.com/chandan-sarma/ISB$\_$interaction}

\bibitem{IMME} E. P. Wigner, in Proceedings of the Robert A.Welch Foundation Conference on Chemical Research, edited by W. O. Milligan, Vol. 1 (Welch Foundation, Houston, 1957), p. 86.

\bibitem{AME2020} M. Wang, W. J. Huang, F. G. Kondev, G. Audi, and S. Naimi, ``The AME 2020 atomic mass evaluation (II). Tables, graphs and references'', \href{https://iopscience.iop.org/article/10.1088/1674-1137/abddaf}{Chinese Phys. C \textbf{45}, 030003 (2020).}

 


\bibitem{ref:ENSDF} Evaluated Nuclear Structure Data File (ENSDF),
 \href{http://www.nndc.bnl.gov/ensdf/}{http://www.nndc.bnl.gov/ensdf/}.


\bibitem{ref:Henderson_18} J. Henderson, G. Hackman, P. Ruotsalainen, S. R. Stroberg, K.D. Launey, J.D. Holt, \textit{et al.}, ``Testing microscopically derived descriptions of nuclear collectivity: Coulomb excitation of $^{22}$Mg'', \href{https://doi.org/10.1016/j.physletb.2018.05.064}{Phys. Lett. B \textbf{782}, 468 (2018)}

\bibitem{brown_iso} B. A. Brown, B. H. Wildenthal, W. Chung, S. E. Massen, M. Bernas, A. M. Bernstein, R. Miskimen, V. R. Brown, and V. A. Madsen, ``Isovector \textit{E2} matrix elements from electromagnetic transitions in the \textit{s-d} shell: Experiment and shell-model calculations'', \href{https://doi.org/10.1103/PhysRevC.26.2247}{Phys. Rev. C \textbf{26}, 2247 (1982).}

\bibitem{SIA1974}
K. Shimizu, M. Ichimura, and A. Arima,
Suzuki,
``Magnetic moments and GT-type $\beta$-decay matrix elements in nuclei with a LS doubly closed shell plus or minus one nucleon",
\href{https://doi.org/10.1016/0375-9474(74)90407-2}{Nucl. Phys. A \textbf{226}, 282 (1974).}

\bibitem{ACS1983}
A. Arima, T. Cheon, K. Shimizu, H, Hyuga, and T. Suzuki,
``Comments on the effect of $\Delta$(1232 MeV)-hole excitation in quenching the Gamow-Teller matrix element",
\href{https://doi.org/10.1016/0370-2693(83)90778-5}{Phys. Lett. B \textbf{122}, 126 (1983).}

\bibitem{KShell} N. Shimizu, T. Mizusaki, Y. Utsuno and Y. Tsunoda, ``Thick-restart block Lanczos method for large-scale shell-model calculations", \href{https://doi.org/10.1016/j.cpc.2019.06.011}{Comput. Phys. Comm. {\bf 244}, 372 (2019).}


		


\end{thebibliography}

\end{document}